\documentclass[a4paper ,12pt]{article}

\usepackage{bbm}
\usepackage{amsmath}
\usepackage{amssymb}
\usepackage{stmaryrd}
\usepackage{graphicx}
\usepackage{amsthm}
\usepackage{natbib}
\usepackage{caption}
\usepackage{fullpage}

\title{A theory of finite deformation magneto-viscoelasticity}
\author{Prashant Saxena, Mokarram Hossain, Paul Steinmann\\ Chair of Applied Mechanics, University of Erlangen-Nuremberg,\\ Egerlandstrasse 5, 91058 Erlangen, Germany}
\date{}

\begin{document}

\maketitle

\begin{abstract}
This paper deals with the mathematical modelling of large strain magneto-viscoelastic deformations. Energy dissipation is assumed to occur both due to the mechanical viscoelastic effects as well as the resistance offered by the material to magnetisation.
Existence of internal damping mechanisms in the body is considered by decomposing the deformation gradient and the magnetic induction into `elastic' and `viscous' parts.
Constitutive laws for material behaviour and evolution equations for the non-equilibrium fields are derived that agree with the laws of thermodynamics. To illustrate the theory the problems of stress relaxation, magnetic field relaxation, time dependent magnetic induction and strain are formulated and solved for a specific form of the constitutive law. The results, that show the effect of several modelling parameters on the deformation and magnetisation process, are illustrated graphically. 
\end{abstract}


\\

Original version published in the \emph{International Journal of Solids and Structures} 50(24): 3886--3897 (2013) doi: 10.1016/j.ijsolstr.2013.07.024


\section{Introduction}

Magnetorheological elastomers (MREs) are materials that change their mechanical behaviour in response to the application of an external magnetic field. 
 These elastomers have received considerable attention in recent years due to their potential uses as variable stiffness actuators for mechanical systems with electronic controls. MREs are particularly useful for their tuneable elastic modulus and a rapid response to the magnetic field, cf. \cite{Bose2012}.
 A common preparation method is mixing magnetically permeable particles into liquid monomer and letting the mixture to polymerise. Curing, when done in the absence of magnetic field, results in an isotropic material while curing in the presence of a magnetic field causes the particles to align in a particular direction and results in a material with a directional anisotropy. The ferromagnetic particles are usually between 1--5 $\mu$m in size and kept between 0--30\% by volume of the entire mixture. Such elastomers have been reported to be prepared and analysed by 
 \cite{Jolly1996}, 
\cite{Ginder2002}, 
\cite{Varga2006}, 
\cite{Boczkowska2009}, and  
\cite{Bose2009}.

 Mathematical modelling of the coupling of electromagnetic fields in deformable materials has beeen an area of active research in the past, see, for example, the works of 
\cite{Pao1978} and 
\cite{Eringen1990}.
 Recently, a new constitutive formulation based on a `total' energy density function has been developed by 
\cite{Dorfmann2003, Dorfmann2004}, wherein the solutions of some boundary value problems were obtained using different energy densities. It has been shown that any one of the magnetic induction vector, magnetic field vector, or the magnetisation vector can be used as an independent variable of the problem and the other two obtained through the constitutive relations.
The relevant equations used by them are based on the classic work of \cite{Pao1978} in which the equations of motion for an isotropic non-polar continuum in an electromagnetic field are described by Maxwell's equations and the mechanical and thermodynamical balance laws.
This formulation has been particularly useful in recent years in dealing with problems related to magnetoelasticity and using this, further boundary value problems on nonlinear deformation and wave propagation have been studied by 
\cite{Bustamante2007a}, 
\cite{Ottenio2008} and 
\cite{Saxena2011}.

The response  to an applied magnetic induction is, however, not exactly instantaneous for all materials. On the application of a sudden external magnetic induction, the magnetic field (or equivalently the magnetisation) developed inside the material is not constant. Starting with some initial non-equilibrium value, it gradually approaches equilibrium in some finite time (say $t_1$) depending on the existing deformation and various material parameters. 
The synthetically developed magnetoelastic materials are usually polymer based, hence also  viscoelastic in nature. Thus there is development of a viscous overstress on deformation or on the application of a body force that vanishes after a time $t_2$ which is usually different from $t_1$ above.
This time-delay in response is a very important factor to consider while designing electromechanical actuators from magnetorheological elastomers. Thus, these two forms of dissiption -- due to mechanical and due to magnetic effects  need to be modelled appropriately.
In order to consider the magnetic and mechanical dissipation effects, the previously stated theory of magnetoelasticity by \cite{Dorfmann2004} is generalised by combining with the existing theory of mechanical viscoelasticity.


Viscoelastic material modelling can generally be classified into two main classes, i.e. purely phenomenologically-motivated and micromechanical based network
models. In some literature, the viscoelastic modelling approach is also divided, on the one hand, due to the nature of the time-dependent part of the stress, on the other
hand, due to the nature of the evolution equation. 
The phenomenological modelling approach can also be distinguished based on the type of internal variables, i.e. stress type internal variables in the form of convolution integrals, cf. 
\cite{Simo1987}, \cite{Holzapfel1996a}, 
\cite{Lion1997}, 
\cite{Kaliske1997}; and strain-type internal variables that originate from a multiplicative decomposition of the deformation gradient, cf. \cite{Reese1998} and 
\cite{Huber2000}.
 The latter group decomposes the deformation gradient into elastic and inelastic parts where the inelastic part is determined from a differential type flow rule. In both the cases, the total stress is decomposed into a viscosity induced overstress and an equilibrium stress that corresponds to stress response at an infinitely slow rate of deformation or the stress response when the time-dependent viscous effects are completely diminished.
 Within the setting of the  multiplicative decomposition of the deformation gradient, 
\cite{Reese1998}  proposed an evolution law which, when linearized around the thermodynamical equilibrium, yields  the finite linear viscoelasatic model of 
\cite{Lubliner1985}. 
\cite{Koprowski-Theiss2011} proposed a nonlinear evolution law, which after being proved to be thermodynamically consistent, has been used in this paper.
For the modelling based on stress-type internal variables, the time-dependent overstress part is expressed as an integral over the deformation history, cf., \cite{Simo1987} and \cite{Amin2006}.

The second class of viscoelastic material modelling is based on micromechanical theories derived using the underlying molecular structures, see, for example, the works of \cite{Bird1987}, \cite{Doi1988}, \cite{Bergstrom1998} and \cite{Miehe2005}. These have been developed over the years to describe the viscous behaviour of molten polymers and physically cross-linked rubber-like materials. The bead-spring model of \cite{Bird1987}, the reptation-type tube models of \cite{DeGennes1971} and  \cite{Doi1988},  and the transient network models of \cite{Green1946} can be mentioned as examples in this area. The theory for transient network models explains the stress relaxation phenomena as a consequence of breakage and reformulation of the polymer
cross-links constantly, cf. 
\cite{Green1946} and 
\cite{Reese2003}. Reptation-type tube models are developed for the definition of the motion of a single chain in a polymer gel. The constraints on the free motion of a single chain are qualitatively modelled as a tube-like constraint and the motion of the chain is described as a combination of Brownian motion within and reptational motion along the tube. Recently, a growing interest can be observed to combine these approaches which yield the so-called micromechanically motivated models, see, for example, \cite{Linder2011}.

As a first step in the magneto-viscoelastic modelling, we model an isotropic material and take a phenomenological approach based on the multiplicative decomposition of the deformation gradient in line with 
\cite{Lubliner1985}.
An additive decomposition of the magnetic induction vector into equilibrium and non-equilibrium parts is  proposed to model magnetic dissipation phenomena. The equilibrium part of the energy is taken to be a generalisation of the Mooney--Rivlin elastic model to include magnetic effects, while the non-equilibrium part is a slightly simplified version that looks like a neo-Hookean type magnetoelastic model. Using a Clausius--Duhem form of the second law of thermodynamics, we obtain evolution equations for these physical quantities to be able to perform numerical calculations.

This paper is organised as follows. In Section 2, the theory of nonlinear magneto-viscoelasticity is presented taking into account the case of finite deformation. Starting with the governing Maxwell's equations and the laws of momentum balance, we show the existence of a total stress tensor of 
\cite{Dorfmann2003}. The deformation gradient and the magnetic induction are decomposed into equilibrium and non-equilibrium parts. Using the laws of thermodynamics and a form of the Helmholtz free energy function, constitutive equations are derived  along with the conditions to be satisfied by the evolution equations of the non-equilibrium quantities.
 
In Section 3, for the purpose of obtaining numerical solutions  the energy density function and the evolution equations for the non-equilibrium quantities are specialised to specific forms. Several material parameters to model magneto-viscoelastic coupling are introduced in this step. 
In Section 4, we consider four different types of deformation and magnetisation processes to study the effects of the underlying magnetic induction, deformation, strain rate and magnetic induction rate on the total Cauchy stress and the magnetic field relaxation process.  
It is observed that changing the newly defined magneto-viscoelastic coupling parameters  can affect the magnitude of the overstress, the excess magnetic field and their decay times. Initial deformation can affect the decay time and magnitude of the induced magnetic field. 
Effects of the deformation, applied magnetic induction and the material parameters on the computed physical quantities (such as stress and magnetic field) are illustrated graphically.
 Section 5 contains some brief concluding remarks.

\section{Theory of nonlinear magneto-viscoelasticity}

We consider an incompressible magnetoelastic material which, when undeformed and unstressed and in the absence of magnetic fields, occupies the material configuration $\mathcal B_0$ with boundary $\partial \mathcal B_0$. It is then subjected to a static deformation due to the combined action of a magnetic field and mechanical surface and body forces.
The spatial configuration at time $t$ is denoted by $\mathcal B_t$ with a boundary $\partial \mathcal B_t$. The two configurations are related by a deformation function $\boldsymbol \chi$ which maps every point $\mathbf X \in \mathcal B_0$ to a point $\mathbf x = \boldsymbol \chi(\mathbf X, t) \in \mathcal B_t$.
The deformation gradient is defined as $\mathbf{F} = \mbox{Grad}\, \boldsymbol \chi$, where Grad is the gradient operator with respect to $\mathbf{X}$. Its determinant is given by $J = \mbox{det} \, \mathbf{F} \equiv 1$ for the present case of incompressibility.

\begin{figure}
\begin{center}
\includegraphics[scale=0.9]{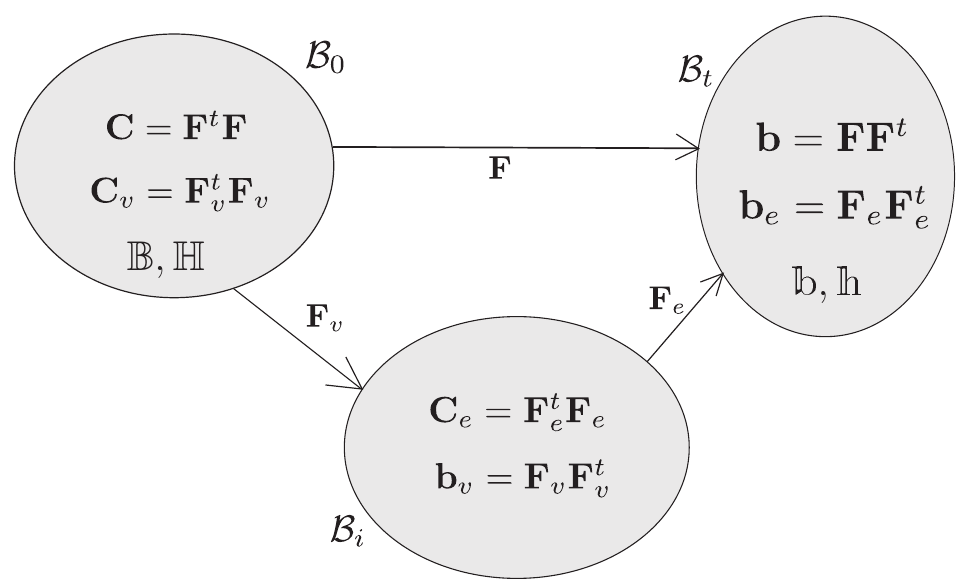}
\end{center}
\caption{The material, intermediate, and spatial configurations with the corresponding magnetic vectors and deformation tensors.}
\label{fig: potatoes}
\end{figure}

To take into account mechanical viscous effects, we assume the existence of an intermediate configuration $\mathcal B_i$ that is related to $\mathcal B_t$ by a purely elastic deformation and is related to $\mathcal B_0$ by a pure viscous motion. The intermediate configuration $\mathcal B_i$ is postulated only to model the dissipation effects. This is in parallel to  the energy-conserving magnetoelastic deformation from $\mathcal B_0$ to $\mathcal B_t$.
Following \cite{Lubliner1985} and \cite{Reese1998}, this motivates the decomposition of the deformation gradient into an elastic and a viscous part as
\begin{equation} \label{F Fe Fv decomposition}
\mathbf{F} = \mathbf{F}_e \mathbf{F}_v.
\end{equation}
For future use we define the right  and the left Cauchy--Green strain tensors as $\mathbf{C} = \mathbf{F}^t \mathbf{F}$ and  $\mathbf{b} = \mathbf{F} \mathbf{F}^t$, respectively. Similar quantities corresponding to $\mathbf{F}_v$ and $\mathbf{F}_e$ are defined in $\mathcal B_0, \mathcal B_i$ and $\mathcal B_t$ as shown in Fig.~\ref{fig: potatoes}.

It is further assumed that the material is electrically non-conducting and there are no electric fields.
Let $\boldsymbol{\sigma}$ be the `mechanical' Cauchy stress tensor and $\boldsymbol \tau$  the total Cauchy stress tensor  (see, for example, 
\cite{Dorfmann2004} for its definition), $\rho$ the mass density, $\mathbf{f}_m$ the mechanical body force per unit mass, $\mathbf{a}$ the acceleration of a point, $\mathbbm{f}$ the electromagnetic body force per unit volume, $\mathbbm{h}$ the magnetic field vector, $\mathbbm{b}$ the magnetic induction vector, and $\mathbbm{m}$ the magnetisation vector. The following balance laws need to be satisfied
\begin{equation} \label{gov eulerian 1}
\mbox{div}\, \boldsymbol{\sigma} + \mathbbm{f} + \rho \mathbf{f}_m = \rho \mathbf{a}, \quad \mbox{div}\, \boldsymbol{\tau} + \rho \mathbf{f}_m = \rho \mathbf{a}, \quad \boldsymbol \tau ^t = \boldsymbol{\tau},
\end{equation}
\begin{equation} \label{gov eulerian 2}
 \mbox{curl}\, \mathbbm{h} = \boldsymbol 0, \quad \mbox{div} \, \mathbbm{b} = 0.
\end{equation}
In Eq.~\eqref{gov eulerian 1}, the first two equations are equivalent forms of the balance of linear momentum and the third is the angular momentum balance equation. Eq.~\eqref{gov eulerian 2}$_1$ is the specialisation of Amp\`{e}re's law appropriate to the present situation and Eq.~\eqref{gov eulerian 2}$_2$ is the statement of impossibility of the existence of magnetic monopoles. Here and henceforth, grad, div, curl denote the standard differential operators in $\mathcal B_t$ while Grad, Div, Curl denote the corresponding operators in $\mathcal B_0$. The three magnetic vectors are connected through the standard relation
\begin{equation} \label{three field relation eulerian}
\mathbbm{b} = \mu_0 \left[ \mathbbm{h} + \mathbbm{m} \right],
\end{equation}
$\mu_0$ being the magnetic permeability of vacuum.
The connection between $\boldsymbol \sigma$ and $ \boldsymbol \tau$ is
\begin{equation} \label{sigma tau rel}
\boldsymbol \tau = \boldsymbol \sigma + \mu_0^{-1} \left[ \mathbbm b \otimes \mathbbm b - \frac{1}{2} \left[ \mathbbm b \cdot \mathbbm b \right] \mathbf{i} \right] + \left[ \mathbbm{m} \cdot \mathbbm{b} \right] \mathbf{i} - \mathbbm{b} \otimes \mathbbm{m},
\end{equation}
where $\mathbf{i}$ is the second order identity tensor in $\mathcal B_t$ and we have used the expression for the magnetic body force as $\mathbbm{f} = \left[ \mbox{grad}\, \mathbbm{b} \right]^t \mathbbm{m}$; see, for example, 
\cite{Pao1978}.

The total Piola-Kirchhoff stress and the Lagrangian forms of $\mathbbm h, \mathbbm m$, and $\mathbbm b$ for an incompressible material ($J=1$) are defined by
\begin{align}
\mathbf{S} &= \mathbf{F}^{-1} \boldsymbol{\tau} \mathbf{F}^{-t}  \nonumber \\
&= \mathbf{F}^{-1} \boldsymbol{\sigma} \mathbf{F}^{-t} + \mu_0^{-1} \left[ \mathbb B\otimes \mathbb B - \frac{1}{2} \left[ \left[ \mathbf{C} \mathbb{B} \right] \cdot \mathbb B \right] \mathbf{C}^{-1} \right] + \left[ \mathbb M \cdot \mathbb B \right] \mathbf{C}^{-1} - \left[ \mathbb B\otimes \mathbb M \right] \mathbf{C}^{-1}, \label{piola kirchoff cauchy rel}
\end{align}
\begin{equation} \label{ref-curr relations}
\mathbbm{H} = \mathbf{F}^t \mathbbm{h}, \quad \mathbb{M} = \mathbf{F}^t \mathbbm{m}, \quad \mathbbm{B} = \mathbf{F}^{-1} \mathbbm{b}.
\end{equation}
We use the above relations to rewrite the governing equations \eqref{gov eulerian 1} and \eqref{gov eulerian 2} in terms of the Lagrangian variables as
\begin{equation} \label{pullback}
\mbox{Div}\left(\mathbf{S F}^t \right) + \rho \mathbf{f}_m = \rho \mathbf{a}, \quad \mathbf{S}^t = \mathbf{S}, \quad \mbox{Curl}\, \mathbbm H = \boldsymbol 0, \quad \mbox{Div}\, \mathbbm B = 0,
\end{equation}
while the relation \eqref{three field relation eulerian} becomes
\begin{equation} \label{three field relation lagrangian}
\mathbf{C} \mathbbm{B} = \mu_0 \left[ \mathbb{H} + \mathbb{M} \right].
\end{equation}

In magnetorheological elastomers, in addition to the mechanical viscoelastic effects, we propose that energy dissipation also occurs due to the resistance offered to magnetisation of the material. On the sudden application of a constant magnetic induction, the magnetic field generated inside the material starts from an initial non-equilibrium value and then evolves to approach an equilibrium value. 
To model these effects, we assume the existence of a dissipation mechanism by including magnetic induction like `elastic' and `viscous' internal variables $\mathbbm{b}_e$ and $\mathbbm{b}_v$, respectively. 
Their Lagrangian counterparts are given by $\mathbbm{B}_e$ and $\mathbbm{B}_v$ such that 
\begin{equation}
\mathbbm{b} = \mathbbm{b}_e + \mathbbm{b}_v, \quad \mathbb{B} = \mathbb{B}_e + \mathbb{B}_v.
\end{equation}
The above additive decomposition of the magnetic induction is motivated by a similar decoupling into elastic and viscous parts of the deformation in viscoelasticity theory. Since magnetic induction is a vector, an additive decomposition is applied here as opposed to the multiplicative decomposition of the deformation gradient in Eq.~\eqref{F Fe Fv decomposition}. 
The behaviour of the internal variables defined in the above is assumed such that if a constant magnetic induction $\mathbb{B}$ is applied at time $t=0$, then at that instant $\mathbb{B}_e = \mathbb{B}$ and $\mathbb{B}_v = \mathbf{0}$. As time progress, the magnetic induction is gradually and entirely transferred to $\mathbb{B}_v$.
 Thus,
\begin{align}
 \mathbb{B}_e = \mathbb{B}, \quad \mathbb{B}_v = \boldsymbol{0} \quad &\mbox{at}\; t=0 ,\nonumber \\
 \mathbb{B}_e \rightarrow \boldsymbol{0}, \quad \mathbb{B}_v \rightarrow \mathbb{B} \quad &\mbox{as}\; t\rightarrow \infty. \label{Bv goes to Be}
\end{align}

\subsection{Thermodynamics and constitutive modelling}
In this section, starting with the laws of thermodynamics, we use the balance equations \eqref{gov eulerian 1} and \eqref{gov eulerian 2} to obtain constitutive laws that define the material behaviour. Necessary conditions of energy dissipation associated with the viscous and the magnetic dissipation processes are obtained that need to be satisfied by any magnetoelastic deformation to be thermodynamically admissible.


Balance of energy is written in the local form as (cf. \cite{Pao1978}, \cite{Dorfmann2003})
\begin{equation}
\rho \frac{d}{dt} \left( U + \frac{1}{2} |\mathbf{v}|^2 \right) + \mbox{div}\, \mathbf{Q} =  \mbox{div} \left( \boldsymbol{\sigma} \mathbf{v} \right) + \left[ \rho \mathbf{f}_m + \mathbbm{f} \right] \cdot \mathbf{v} + \rho R + \mathbbm{w}_e,
\end{equation}
where $U$ and $R$ denote the internal energy and radiant heating per unit mass, $\mathbf{Q}$ is the heat flux, $\boldsymbol{\sigma}$ is the purely mechanical Cauchy stress, 
$\mathbf{f}_m$ is the mechanical body force (assumed to be zero in the analysis later), and $\mathbbm{w}_e$ is the electromagnetic power given for the present case as $\mathbbm w_e = - \mathbbm{m} \cdot d \mathbbm{b}/dt$.

Let $S$ be the specific entropy and $\vartheta$ be the temperature. On introducing a specific Helmholtz free energy $\Psi$ through
\begin{equation}
\Psi = U - \vartheta S,
\end{equation}
and using the Clausius--Duhem form of the second law of Thermodynamics
\begin{equation}
\rho \frac{dS}{dt} + \mbox{div} \left( \frac{\mathbf{Q}}{\vartheta} \right) - \rho \frac{R}{\vartheta} \ge 0,
\end{equation}
we arrive at the following inequality
\begin{equation} \label{SLT 1}
- \rho \frac{d \Psi}{dt} + \mathbf{F}^{-1} \boldsymbol{\sigma} \colon \frac{d \mathbf{F}}{dt} - \mathbbm{m} \cdot \frac{d \mathbbm{b}}{dt}  \ge 0.
\end{equation}
The symbol $:$ denotes a double contraction operation between two second order tensors. In the calculations above use has been made of Eq.~\eqref{gov eulerian 1}$_1$ and the temperature is assumed to be constant.

We now introduce a total energy function similar to the one used by 
\cite{Dorfmann2004}
\begin{equation}
\Omega(\mathbf{F}, \mathbf{C}_v, \mathbb{B}, \mathbb{B}_v) = \rho \Psi (\mathbf{C}, \mathbf{C}_v, \mathbf{F} \mathbb{B}, \mathbf{F} \mathbb{B}_v) + \frac{1}{2 \mu_0} \mathbb{B} \cdot \left[ \mathbf{C} \mathbb{B} \right].
\end{equation}
This considers the magnetic induction vector $\mathbb{B}$ as an independent quantity and leaves the magnetic field $\mathbb{H}$ to be determined using a constitutive law. The magnetisation $\mathbb{M}$, if required, can be obtained using the relation \eqref{three field relation lagrangian}.

The above equation, on differentiation with respect to time gives
\begin{equation} \label{eq. 17}
-\rho \frac{d \Psi}{dt} = - \frac{d \Omega}{dt} + \frac{1}{2 \mu_0} \mathbb B \cdot \left[ \frac{d \mathbf{C}}{dt} \mathbb B \right] + \frac{1}{\mu_0} \left[ \mathbf{C} \mathbb B\right] \cdot \frac{d \mathbb B}{dt},
\end{equation}
while using \eqref{ref-curr relations}, we obtain
\begin{equation} \label{eq. 18}
- \mathbbm m \cdot \frac{d \mathbbm b}{dt} = - \mathbb M \cdot \left[ \mathbf{F}^{-1} \frac{d \mathbf{F}}{dt} \mathbb B \right] - \mathbb M \cdot \frac{d \mathbb B}{dt}.
\end{equation}

Note that the constraint of incompressibility can be expressed as
\begin{equation}
\frac{d J^2}{dt} = \mathbf{C}^{-1} : \frac{d \mathbf{C}}{dt} = 0.
\end{equation}
For an incompressible material, $ d \mathbf{C}/dt$ is not arbitrary, but the inequality \eqref{SLT 1} must be satisfied for every $d \mathbf{C}/dt$ governed by the above constraint. Consequently adding a scalar multiple of this zero term  and substituting Eqs. \eqref{eq. 17} and \eqref{eq. 18} to the inequality  \eqref{SLT 1}, a form of the second law of  thermodynamics is obtained in terms of the `total' Piola-Kirchhoff stress tensor and the physical quantities defined in their Lagrangian description as
\begin{equation} \label{SLT 2}
- \frac{d \Omega}{dt} + \frac{1}{2} \left[ \mathbf{S} + p \mathbf{C}^{-1} \right]: \frac{d \mathbf{C}}{dt}  + \mathbb{H} \cdot \frac{d \mathbb{B}}{dt} \ge 0,
\end{equation}
$p$ being a Lagrange multiplier associated with the constraint.
It is noted here that one can equivalently use a nominal stress tensor $\mathbf{T} = \mathbf{F}^{-1} \boldsymbol{\tau}$ instead of $\mathbf{S}$ and take $\mathbf{F}$ as an independent variable instead of $\mathbf{C}$ and substitute in \eqref{SLT 1}. This would result in an inequality  that yields the constitutive relations used by \cite{Dorfmann2004}.

Taking partial derivatives of $\Omega$ with respect to its arguments and substituting in the inequality above, we get
\begin{align}
\frac{1}{2}\left[ \mathbf{S} - 2 \frac{\partial \Omega}{\partial \mathbf{C}} + p\mathbf{C}^{-1} \right] : \frac{ d \mathbf{C}}{dt} + \left[ \mathbb{H} - \frac{\partial \Omega}{\partial \mathbb{B}} \right] \cdot \frac{d \mathbb{B}}{dt} \nonumber \\
 - \frac{\partial \Omega}{\partial \mathbf{C}_v} : \frac{d \mathbf{C}_v}{dt} - \frac{\partial \Omega}{\partial \mathbb{B}_v} \cdot \frac{d \mathbb{B}_v}{dt} \ge 0,
\end{align}

From the arguments of \cite{Coleman1963},
the following constitutive equations are obtained
\begin{equation}
\mathbf{S} = 2 \frac{\partial \Omega}{\partial \mathbf{C}} - p\mathbf{C}^{-1}, \quad \mathbb{H} = \frac{\partial \Omega}{\partial \mathbb{B}},
\end{equation}
along with the dissipation condition
\begin{equation} 
\frac{\partial \Omega}{\partial \mathbf{C}_v} \colon \frac{d \mathbf{C}_v}{dt} + \frac{\partial \Omega}{\partial \mathbb{B}_v} \cdot \frac{d \mathbb{B}_v}{dt} \le 0.
\end{equation}

For the sake of simplicity in the computations later, we further assume that the non-equilibrium magnetic induction $\mathbb B_v$ and the non-equilibrium strain tensor $\mathbf{C}_v$ are independent from each other. This reduces the above inequality to the following  separate conditions
\begin{equation} \label{Ce Be dissipation 1}
  \frac{\partial \Omega}{\partial \mathbf{C}_v} \colon \frac{d \mathbf{C}_v}{dt} \le 0, \quad \frac{\partial \Omega}{\partial \mathbb{B}_v} \cdot \frac{d \mathbb{B}_v}{dt} \le 0,
\end{equation}
which should be satisfied by any magnetoelastic deformation process to be thermodynamically admissible.

The total  energy stored in the body can be split into an equilibrium part associated with the direct deformation from $\mathcal B_0$ to $\mathcal B_t$, and a viscous part due to the internal variable $\mathbb B_e$ and the elastic deformation from $\mathcal B_i$ to  $\mathcal B_t$. This is a slightly general form of the purely mechanical energy decomposition by
 \cite{Reese1998}
\begin{equation} \label{energy split}
\Omega \left(\mathbf{C}, \mathbf{C}_v, \mathbb{B}, \mathbb{B}_v \right) = \Omega_e (\mathbf{C}, \mathbb{B}) + \Omega_v \left( \mathbf{C}, \mathbf{C}_v, \mathbb{B}, \mathbb{B}_v \right).
\end{equation}
Here, the viscous part of the energy depends on the viscous parts of the deformation and the magnetic induction. Thus the arguments of $\Omega_v$ can be equivalently changed as either one of $\Omega_v \left( \mathbf{C}, \mathbf{C}_v, \mathbb{B}_e \right)$, $\Omega_v \left( \mathbf{C}_e, \mathbb{B}, \mathbb{B}_v \right)$ or $\Omega_v \left( \mathbf{C}_e, \mathbb{B}_e \right)$.

Substituting this form of $\Omega$ into the inequalities \eqref{Ce Be dissipation 1} we obtain the dissipation conditions that should be necessarily met in order to satisfy the second law of thermodynamics
\begin{equation} \label{Ce thermo constraint}
 \frac{\partial \Omega_v}{\partial \mathbf{C}_v} : \frac{d \mathbf{C}_v}{dt} \le 0,
\end{equation}
\begin{equation} \label{Bv thermo constraint}
 \frac{\partial \Omega_v}{\partial \mathbb{B}_v} \cdot \frac{d \mathbb{B}_v}{dt} \le 0.
\end{equation}


It is noted here that the above theory can easily be generalised to include multiple dissipation mechanisms in the body. 
In the case of $M$ mechanical and $N$ magnetic mechanisms, we may define $\mathbf{F}_e^1,...,\mathbf{F}_e^M; \mathbf{F}_v^1, ...,\mathbf{F}_v^M $; $\mathbb{B}_e^1, ...,\mathbb{B}_e^N $; $\mathbb{B}_v^1,...,\mathbb{B}_v^N$ such that
\begin{align}
\mathbf{F} = \mathbf{F}_e^i \mathbf{F}_v^i, \quad \forall i=1,...,M, \\
\mathbb{B} = \mathbb{B}_e^j + \mathbb{B}_v^j, \quad \forall j=1,...,N.
\end{align}

The dissipation condition to be satisfied in this general case is
\begin{equation}
\sum\limits_{i=1}^M \frac{\partial \Omega}{\partial \mathbf{C}_v^i} \colon \frac{d \mathbf{C}_v^i}{dt} + \sum\limits_{j=1}^N \frac{\partial \Omega}{\partial \mathbb{B}_v^j} \cdot \frac{d \mathbb{B}_v^j}{dt} \le 0.
\end{equation}

\section{Specialised constitutive laws}
With a motivation of obtaining numerical solutions to some magneto-viscoelastic deformation problems, we specialise the energy in \eqref{energy split} to specific forms in this section. Evolution equations for $\mathbf{C}_v$ and $\mathbb B_v$ are also derived that satisfy the thermodynamic constraints \eqref{Ce thermo constraint} and \eqref{Bv thermo constraint}.


\subsection{Energy functions}
The material is assumed to be isotropic following which the equilibrium part of the energy density function is considered to be a generalisation of the classical Mooney--Rivlin function to magnetoelasticity of the form 
\begin{align} 
\Omega_e = \frac{\mu_e}{4} \left[ 1 + \alpha_e \, \mbox{tanh} \left(\frac{I_4}{m_e} \right) \right] \big[ \left[ 1+n \right]\left[ I_1-3 \right] + \left[ 1-n \right]\left[ I_2-3 \right] \big] \nonumber  \\
 + q I_4 + r I_6 , \label{eqbm energy function}
\end{align}
where $I_1, I_2, I_4$ and $I_6$ are the standard scalar invariants in magnetoelasticity (see, for example, 
\cite{Dorfmann2004}) defined as
\begin{align} 
I_1 = \mathbf{C} : \mathbf{I}, \quad I_2 = \frac{1}{2} \left[I_1^2 -  \mathbf{C}^2 : \mathbf{I} \right], \quad I_4 = [\mathbb{B} \otimes \mathbb{B}] : \mathbf{I}, \nonumber \\
 I_6 = \big[ \left[\mathbf{C} \mathbb{B} \right] \otimes \left[\mathbf{C} \mathbb{B}\right] \big] : \mathbf{I}, \label{invariant definition}
\end{align}
$\mathbf{I}$ being the second order identity tensor in $\mathcal B_0$.

This is a slight generalisation of a  Mooney--Rivlin type magnetoelastic energy function proposed by 
 \cite{Ottenio2008}. Here $\mu_e$ is the shear modulus of the material in the absence of a magnetic field and $n$ is a dimensionless parameter restricted to the range $-1 \le n \le 1 $, as for the classical Mooney--Rivlin model. The term $\left[ 1 + \alpha_e \mbox{tanh} \left(I_4/m_e \right) \right]$ corresponds to an increase in the stiffness due to magnetisation and the phenomenon of magnetic saturation after a critical value of magnetisation. The parameter $m_e$ is required for the purpose of non-dimensionalisation while $\alpha_e$ is a dimensionless positive parameter for scaling. The magnetoelastic coupling parameters $q$ and $r$ have the dimensions of $\mu_0^{-1}$. For $\alpha_e=q=r=0$, this simplifies to the classical Mooney--Rivlin elastic energy density function widely used to model elastomers.

Let the natural basis vectors in $\mathcal B_0$ be identified with a set of covariant basis vectors $\{ \mathbf{G}_\alpha \}$ and its dual basis with a set of contravariant basis vectors $\{ \mathbf{G}^\alpha \}$, $\alpha \in \{1,2,3\}$. Similarly defining $\{ \mathbf{g}_\alpha \}$ and $\{ \mathbf{g}^\alpha \}$  for $\mathcal B_t$ and $\{ \bar{\mathbf{g}}_\alpha \}$ and $\{ \bar{\mathbf{g}}^\alpha \}$  for $\mathcal B_i$, we obtain the following forms for the deformation gradient and the right Cauchy--Green strain tensors
\begin{equation}
\mathbf{F} = \mathbf{g}_\alpha \otimes \mathbf{G}^\alpha , \quad \mathbf{C} = \left[ \mathbf{g}_\alpha \cdot \mathbf{g}_{\beta} \right] \mathbf{G}^{\alpha} \otimes \mathbf{G}^{\beta}, \quad \mathbf{C}_v = \left[ \bar{\mathbf{g}}_\alpha \cdot \bar{\mathbf{g}}_{\beta} \right] \mathbf{G}^{\alpha} \otimes \mathbf{G}^{\beta}.
\end{equation}

If a vector $\mathbb B$ is written in the natural covariant basis as $\mathbb B = B^\alpha \mathbf{G}_\alpha$, then $\mathbf{C} \mathbb B = B^\beta \left[\mathbf{g}_\alpha \cdot \mathbf{g}_\beta \right] \mathbf{G}^\alpha$. Thus the identity tensor used for double contraction in \eqref{invariant definition}$_{1,4}$ needs to have a covariant set of basis vectors while that used in \eqref{invariant definition}$_3$ requires a contravariant basis.

To obtain the non-equilibrium part of the energy density function, we consider a simplification of the Mooney--Rivlin type energy in \eqref{eqbm energy function} by taking $\alpha_e=0$ and $n=1$.
Since the non-equilibrium  part of energy should depend only on the elastic parts of the deformation gradient and the magnetic induction as assumed earlier, we require $I_1$ in \eqref{invariant definition}$_1$ to obtain the value $\mathbf{C}_e : \mathbf{I}$. This is equivalent to the expression $\mathbf{C}: \mathbf{C}_v^{-1}$ on using Eq.~\eqref{F Fe Fv decomposition}.
Hence to obtain the non-equilibrium energy, instead of a double contraction with the identity tensor to obtain the invariants in \eqref{invariant definition}, we do so by the contravariant tensor $\mathbf{C}_v$ and the covariant tensor $\mathbf{C}_v^{-1}$ as appropriate and replace the Lagrangian vector $\mathbb B$  by $\mathbb B_e$. The energy function thus obtained is  similar to a generalisation of  a neo-Hookean type energy to include magnetic effects
\begin{align} 
\Omega_v( \mathbf{C},  \mathbf{C}_v,  \mathbb B, \mathbb B_v) = \frac{\mu_v}{2} \left[ \mathbf{C} : \mathbf{C}_v^{-1} - 3 \right]  
 + q_v \big[ \left[ \mathbb B - \mathbb B_v \right] \otimes \left[ \mathbb B - \mathbb B_v \right] \big] : \mathbf{C}_v \nonumber \\
 + r_v \Big[ \big[\mathbf{C}\left[\mathbb B - \mathbb B_v \right] \big] \otimes \big[\mathbf{C}\left[\mathbb B - \mathbb B_v \right] \big] \Big] : \mathbf{C}_v^{-1}.  \label{dissipation energy function}
\end{align}
The magneto-viscoelastic coupling parameters $q_v$ and $r_v$ here are similar to $q$ and $r$ used in the definition of $\Omega_e$ and we have used the relation $\mathbb B_e = \mathbb B - \mathbb B_v$.

A strong coupling between the mechanical viscous measure $\mathbf{C}_v$ and the magnetic non-equilibrium quantity $\mathbb B_v$ is noted here. For the sake of simplicity of our calculations and in the absence of any experimental data, we simplify the above expression by assuming that the magnetic non-equilibrium effects are coupled only with the total deformation $\mathbf{C}$ and not with its viscous component $\mathbf{C}_v$. This assumption was also used earlier to arrive at the dissipation conditions \eqref{Ce Be dissipation 1}. Thus, by replacing $\mathbf{C}_v$ with $\mathbf{I}$ in \eqref{dissipation energy function}, a simpler form of the non-equilibrium energy is obtained as
\begin{align}
\Omega_v( \mathbf{C}, \mathbf{C}_v,  \mathbb B, \mathbb B_v) 
= \frac{\mu_v}{2} \left[ \mathbf{C}_v^{-1} : \mathbf{C} - 3 \right] + q_v \big[ \left[\mathbb B - \mathbb B_v \right] \otimes \left[\mathbb B - \mathbb B_v \right] \big] : \mathbf{I} \nonumber \\
 + r_v \, \Big[ \big[\mathbf{C}\left[\mathbb B - \mathbb B_v \right] \big] \otimes \big[\mathbf{C} \left[\mathbb B - \mathbb B_v \right] \big] \Big] : \mathbf{I}.  \label{dissipation energy function 2}
\end{align}

\subsection{Evolution equations}

In order to completely define the magneto-viscoelastic behaviour of a solid material, along with the balance laws and the energy density functions defined in the previous sub-section, we also require evolution laws for the `viscous' quantities $\mathbb B_v$ and $\mathbf{C}_v$. These are postulated such that the laws of thermodynamics are satisfied at every instant and $\mathbb B_v$ and $\mathbf{C}_v$ stop evolving when the system reaches an equilibrium state.

For the non-equilibrium part of the magnetic induction we consider the following evolution equation such that the left side of inequality \eqref{Bv thermo constraint} becomes a negative semi-definite quadratic form, automatically satisfying the thermodynamic constraint. Thus,
\begin{align}
 \frac{d \mathbb{B}_v}{dt} & = - \frac{\mu_0}{T_m} \frac{\partial \Omega_v}{\partial \mathbb{B}_v}, \nonumber \\ 
& =  \frac{2 \mu_0}{T_m} \left[ q_v \mathbf{I} + r_v \mathbf{C} \right] \left[ \mathbb B - \mathbb B_v \right]. \label{Bv evolution equation}
\end{align}

For the mechanical viscous strain tensor, we use the evolution equation as proposed by 
\cite{Koprowski-Theiss2011}
\begin{equation} \label{mechanical evolution}
\frac{d \mathbf{C}_v}{dt} = \frac{1}{T_v} \left[ \mathbf{C} - \frac{1}{3}  \left[ \mathbf{C} : \mathbf{C}_v^{-1} \right] \mathbf{C}_v \right].
\end{equation}
In the equations above, $T_v$ is the specific relaxation time for the viscoelastic component of the dissipation mechanism while $T_m$ is the specific relaxation time for its magnetic component. Typically $T_v$ is of the order of some minutes or even upto a few hours  while $T_m$ is of the order of a few seconds or some milliseconds.

It now remains to prove that the evolution equation \eqref{mechanical evolution} is thermodynamically consistent with the energy density function \eqref{dissipation energy function 2}, i.e. they satisfy the constraint \eqref{Ce thermo constraint}.

Consider a fourth order projection tensor defined as
\begin{equation}
 \mathbb{I}_{\mathbf{C}_v}^{\mbox{\scriptsize dev}} = \mathbb{I} - \frac{1}{3} \mathbf{C}_v \otimes \mathbf{C}_v^{-1},
\end{equation}
where $\mathbb I$ is the fourth order symmetric identity tensor given in component form as
\begin{equation}
\left[ \mathbb I \right]_{ijkl} = \frac{1}{2} \left( \delta_{ik} \delta_{jl} + \delta_{il} \delta_{jk} \right),
\end{equation}
$\delta_{ij}$ being the Kronecker-Delta.
On a double contraction of this tensor $\mathbb{I}_{\mathbf{C}_v}^{\mbox{\scriptsize dev}}$ with $\mathbf{C}$, a multiple of the right side of the evolution law \eqref{mechanical evolution} is obtained. Thus,
\begin{equation} \label{eq 37}
\frac{1}{T_v} \mathbb I_{\mathbf{C}_v}^{\mbox{\scriptsize dev}} : \mathbf{C} = \frac{d \mathbf{C}_v}{dt}.
\end{equation}

From Eq.~\eqref{dissipation energy function 2}, we obtain
\begin{equation}
\frac{\partial  \Omega_v}{\partial  \mathbf{C}_v} = - \frac{\mu_v}{2} \mathbf{C}_v^{-1} \mathbf{C} \mathbf{C}_v^{-1} = \frac{\mu_v}{2} \mathbf{C} : \mathbb{I}_{\mathbf{C}_v^{-1}},
\end{equation}
where we have used 
the negative-definite fourth order projection tensor $\mathbb I_{\mathbf{C}_v^{-1}}$ which when expanded in component form, gives
\begin{equation}
\left[ \mathbb I_{\mathbf{C}_v^{-1}} \right]_{ijkl} = - \frac{1}{2} \left[ \left[ \mathbf{C}_v^{-1} \right]_{ik} \left[ \mathbf{C}_v^{-1} \right]_{jl} + \left[ \mathbf{C}_v^{-1} \right]_{il} \left[ \mathbf{C}_v^{-1} \right]_{jk} \right].
\end{equation}

Consider the following operation
\begin{equation}
\frac{\partial \Omega_v}{\partial \mathbf{C}_v} : \frac{d \mathbf{C}_v}{dt} = \frac{\mu_v}{2 T_v} \left[ \mathbf{C}: \mathbb{I}_{\mathbf{C}_v^{-1}} \right] : \left[ \mathbb{I}_{\mathbf{C}_v}^{\mbox{\scriptsize dev}}  : \mathbf{C} \right].
\end{equation}

\noindent Since the tensor $\mathbb{I}_{\mathbf{C}_v}^{\mbox{\scriptsize dev}}$ is idempotent, the above expression can be rewritten as 
\begin{align}
\frac{\partial \Omega_v}{\partial \mathbf{C}_v} : \frac{d \mathbf{C}_v}{dt} & = \frac{\mu_v}{2 T_v} \left[ \mathbf{C}: \mathbb{I}_{\mathbf{C}_v^{-1}} \right] : \mathbb{I}_{\mathbf{C}_v}^{\mbox{\scriptsize dev}} : \left[ \mathbb{I}_{\mathbf{C}_v}^{\mbox{\scriptsize dev}}  : \mathbf{C} \right], \\
& = \frac{\mu_v}{2 T_v} \left[ \mathbb{I}_{\mathbf{C}_v}^{\mbox{\scriptsize dev}} : \mathbf{C} \right] :  \mathbb{I}_{\mathbf{C}_v^{-1}} : \left[ \mathbb{I}_{\mathbf{C}_v}^{\mbox{\scriptsize dev}}  : \mathbf{C} \right] \le 0.
\end{align}
The above inequality holds since $\mathbb{I}_{\mathbf{C}_v^{-1}}$ is negative definite.
This is the statement of the dissipation condition \eqref{Ce thermo constraint}, hence the evolution equation \eqref{mechanical evolution} is thermodynamically consistent with the energy density function \eqref{dissipation energy function 2}.

\subsection{Stress and magnetic field calculations}

For the energy functions defined in Eqs.~\eqref{eqbm energy function} and \eqref{dissipation energy function 2}, the Piola--Kirchhoff stress is given as
\begin{equation} \label{Piola-Kirchhof Oe Ov p}
 \mathbf{S} = 2\frac{\partial \Omega_e}{\partial \mathbf{C}}  + 2\frac{\partial \Omega_v}{\partial \mathbf{C}} - p \mathbf{C}^{-1} = \mathbf{S}_e + \mathbf{S}_v - p \mathbf{C}^{-1},
\end{equation}
where
\begin{align} 
 \mathbf{S}_e =  \frac{\mu_e}{2} \left[ 1 + \alpha_e \, \mbox{tanh} \left( \frac{I_4}{m_e} \right) \right] \big[ \left[1+n \right] \mathbf{I} + \left[1-n \right] \left[ I_1 \mathbf{I} - \mathbf{C} \right] \big] \nonumber \\
 + 2 r  \mathbb{B} \otimes \left[\mathbf{C} \mathbb{B} \right] + 2 r \left[ \mathbf{C} \mathbb{B} \right] \otimes \mathbb{B}, \label{Se expression}
\end{align}
and
\begin{equation} \label{Sv expression}
 \mathbf{S}_v = \mu_v \mathbf{C}_v^{-1} + 2 r_v \mathbb{B}_e \otimes \left[ \mathbf{C} \mathbb{B}_e \right] + 2 r_v \left[ \mathbf{C} \mathbb{B}_e  \right] \otimes  \mathbb{B}_e .
\end{equation}


The Lagrangian magnetic field $\mathbb{H}$ is given as
\begin{equation}
 \mathbb{H} = \frac{\partial \Omega_e}{\partial \mathbb{B}} + \frac{\partial \Omega_v}{\partial \mathbb{B}} = \mathbb{H}_e + \mathbb{H}_v,
\end{equation}
where
\begin{align} 
 \mathbb{H}_e = \frac{\mu_e}{2 m_e} \left[ 1 - \alpha_e \, \mbox{tanh}^2 \left(\frac{I_4}{m_e} \right) \right] \big[ \left[ 1+n \right]\left[ I_1-3 \right] + \left[ 1-n \right]\left[ I_2-3 \right] \big] \mathbb{B} \nonumber \\
 + 2 q \mathbb{B} + 2r \mathbf{C}^2 \mathbb{B}, \label{He expression}
\end{align}
and
\begin{equation} \label{Hv final formula}
 \mathbb{H}_v =  2 q_v \mathbb{B}_e + 2 r_v \mathbf{C} \mathbb{B}_e .
\end{equation}
The `viscous' or non-equilibrium magnetic field defined above tends to zero at equilibrium when $\mathbb{B}_e \rightarrow \mathbf{0}$.
Eulerian expressions for the equilibrium values of the total Cauchy stress $\boldsymbol{\tau}$ and the magnetic field $\mathbbm{h}$ can be written using Eqs.~\eqref{Se expression} and \eqref{He expression} as
\begin{align}
\boldsymbol{\tau}_e = \frac{\mu_e}{2} \left[ 1 + \alpha_e \, \mbox{tanh} \left( \frac{I_4}{m_e} \right) \right] \Big[ \left[1+n \right] \mathbf{b} + \left[1-n \right] \left[ I_1 \mathbf{b} - \mathbf{b}^2 \right]\Big] \nonumber \\
 +  r \, \mathbbm{b} \otimes \left[ \mathbf{b} \mathbbm{b} \right] + r \left[ \mathbf{b} \mathbbm{b} \right] \otimes  \mathbbm{b}  ,  \label{taue expression}
\end{align}
\begin{align} \label{he expression}
\mathbbm{h}_e = \frac{\mu_e}{2 m_e} \left[ 1 - \alpha_e \, \mbox{tanh}^2 \left(\frac{I_4}{m_e} \right) \right] \big[ \left[ 1+n \right]\left[ I_1-3 \right]  \nonumber \\
+ \left[ 1-n \right]\left[ I_2-3 \right] \big] \mathbf{b}^{-1} \mathbbm{b}  + 2 q \mathbf{b}^{-1} \mathbbm{b} + 2r \mathbf{b} \mathbbm{b}.
\end{align}

In the case of no deformation ($\mathbf{b = i}$), if $r=0$, then the total equilibrium stress in Eq.~\eqref{taue expression} is unaffected by the magnetic induction and if $q=r=0$ then the equilibrium magnetic field in Eq.~\eqref{he expression} is unaffected by the underlying deformation. 
The coupling caused by the parameters $q$ and $r$ between deformation and magnetic field is inverse to each other. The former causes a directly proportional relation of $\mathbf{b}^{-1}$ to $\mathbbm{h}_e$ while the latter links $\mathbf{b}$ to $\mathbbm{h}_e$.
Similarly $q_v$ and  $r_v$ are required for including these two-way coupling effects for the non-equilibrium quantities.

For the case of no deformation of an isotropic material, the magnetic field should be in the direction of the applied magnetic induction and be directly proportional to the latter.
From Eq.~\eqref{he expression}, this imposes the constraint
\begin{equation} \label{q r constraint}
q+r >0.
\end{equation}

If  $r>0$,  the material stiffens in the direction of the applied magnetic induction while if $\alpha_e>0$, the total stiffness of the material increases isotropically. Both these effects have been observed to be the case in many MREs, see, for example, the results of 
\cite{Jolly1996} and \cite{Varga2006}. However, this need not necessarily be true in general for all magnetoelastic materials and $r$ can have negative values. In this situation and the case of a material with weak magnetoelastic coupling, i.e. very small values of $r$; we require $q>0$ to satisfy the constraint \eqref{q r constraint}.

\section{Numerical examples}
In this section, we model four different types of experiments and obtain the corresponding solutions numerically. The following numerical values of the material parameters are used unless otherwise stated to have a different value for individual computations
\begin{align}
 \mu_0 = 4 \pi \times 10^{-7} \, \mbox{N/A}^2, \quad \mu_e = 2.6 \times 10^5 \, \mbox{N/m}^2, \quad \mu_v = 5\times 10^5 \, \mbox{N/m}^2, \nonumber \\
\alpha_e = 0.3, \quad m_e = 1 \, \mbox{T}^2, \quad n=0.3, \quad q=r=r_v = 1/\mu_0, \quad q_v = 5/\mu_0.
\end{align}
The value of $\mu_e$ is taken to be the value of shear modulus at zero magnetic field for an elastomer filled with 10\% by volume of iron particles, cf. \cite{Jolly1996}. Values of $n,q$ and $r$ are what have been used by \cite{Ottenio2008} and \cite{Saxena2011}. Values of $\mu_v, \alpha_e, q_v, r_v$ are within reasonable physical assumptions and we analyse the dependence of our solutions on the values of these parameters. 

For computations in the following subsections, the equations derived earlier are specialised to a uniaxial deformation and magnetisation in cartesian coordinates. The time-integration is performed using a standard solver \texttt{ode45} from Matlab that employs an explicit Runge--Kutta scheme, cf. \cite{Shampine1997}.

\subsection{Magnetic induction with no deformation}
\begin{figure}
\begin{center}
\includegraphics[scale=0.9]{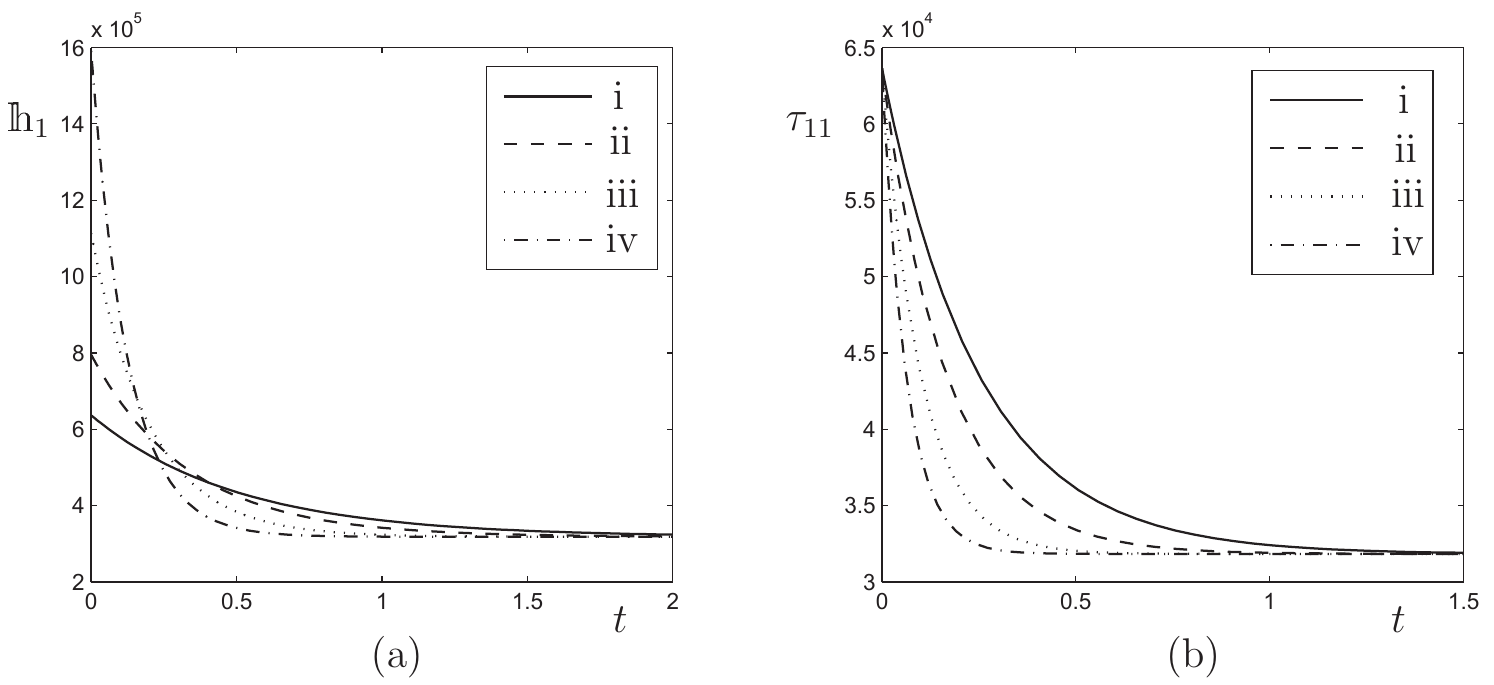}
\end{center}
\caption{Variation of \textbf{(a)} the total magnetic field $\mathbbm{h}_1$ (A/m) and \textbf{(b)} the principal total Cauchy stress $ \tau_{11}$ (N/m$^2$) with time $t$ (s) for no deformation in the presence of a step magnetic induction $\mathbb B_1 = 0.1$ T. Four curves correspond to different values of $q_v$ (i)~$q_v = 1/\mu_0$,  (ii)~$q_v = 2/\mu_0$, (iii)~$q_v = 4/\mu_0$, (iv)~$q_v = 7/\mu_0$.}
\label{fig: no deformation vary_qv}
\end{figure}

\begin{figure}
\begin{center}
\includegraphics[scale=0.9]{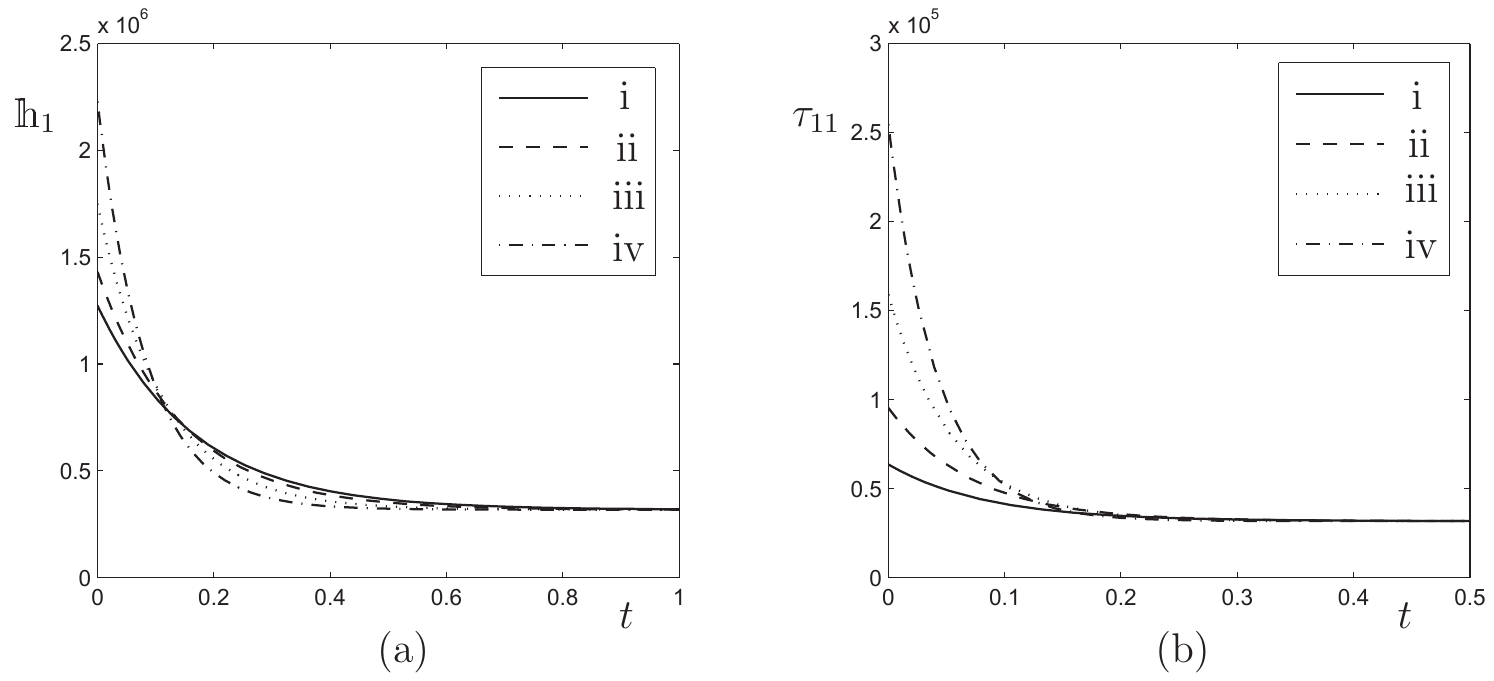}
\end{center}
\caption{Variation of \textbf{(a)} the total magnetic field $\mathbbm{h}_1$ (A/m) and \textbf{(b)} the principal total Cauchy stress $ \tau_{11}$ (N/m$^2$) with time $t$ (s) for no deformation in the presence of a step magnetic induction $\mathbb B_1 = 0.1$ T. Four curves correspond to different values of $r_v$ (i)~$r_v = 1/\mu_0$,  (ii)~$r_v = 2/\mu_0$, (iii)~$r_v = 4/\mu_0$, (iv)~$r_v = 7/\mu_0$.}
\label{fig: no deformation vary_rv}
\end{figure}

\begin{figure}
\begin{center}
\includegraphics[scale=0.9]{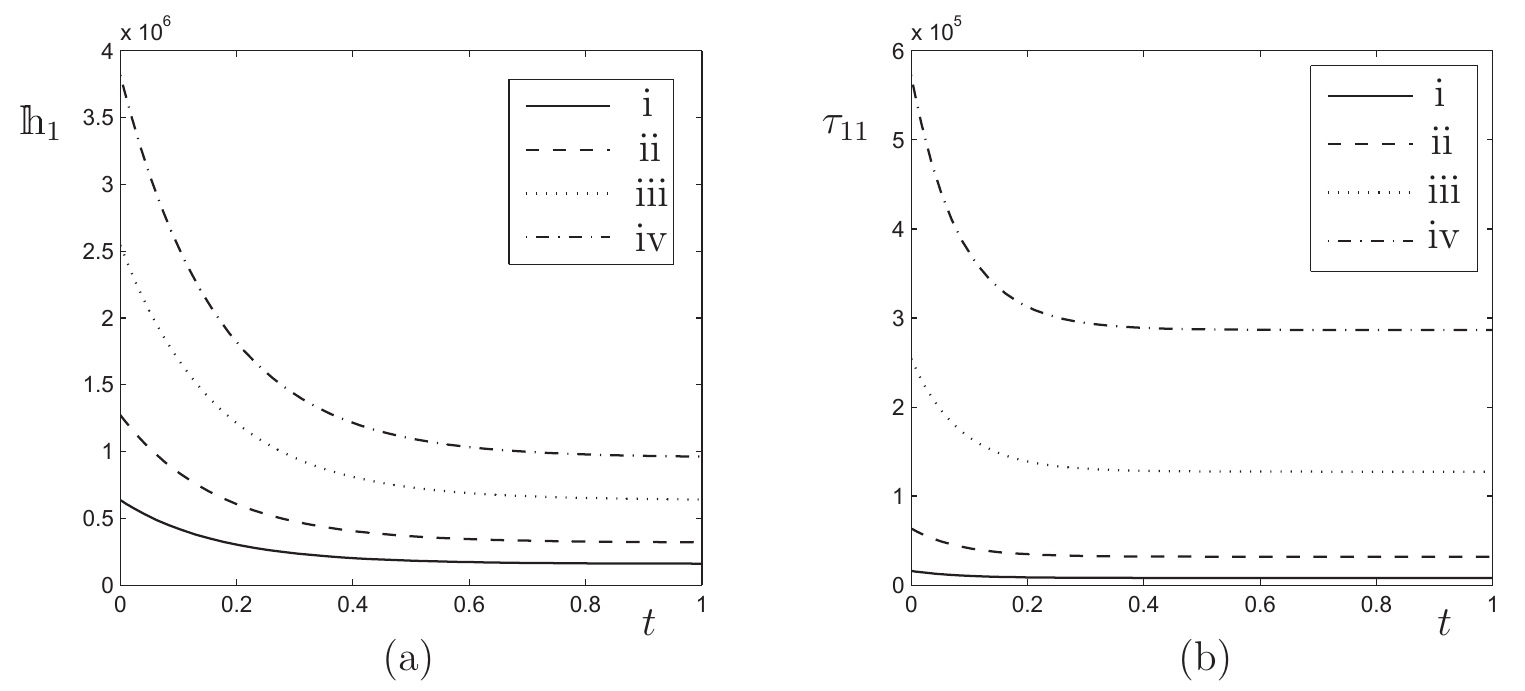}
\end{center}
\caption{Variation of \textbf{(a)} the total magnetic field $\mathbbm{h}_1$ (A/m) and \textbf{(b)} the principal total Cauchy stress $ \tau_{11}$ (N/m$^2$) with time $t$ (s) for different values of the magnetic induction $\mathbb B_1$  (i)~$\mathbb B_1 = 0.05$~T, (ii)~$ \mathbb B_1 = 0.1$~T, (iii)~$\mathbb B_1 = 0.2$~T, (iv)~$ \mathbb B_1 = 0.3$~T.}
\label{fig: no deformation vary_mag}
\end{figure}

With a motivation to isolate and understand the effects of the applied magnetic induction on the magneto-viscoelastic deformation process, we consider no deformation in this first case.
Consider an experiment with the sample held fixed at zero deformation ($\lambda_1 = \lambda_2 = \lambda_3 = 1$) and a sudden but constant magnetic induction applied at time $t=0$.
This results in the generation of a viscous overstress and a temporary increment in the magnetic field, both of which settle down to equilibrium values with time. Variations of the total magnetic field $\mathbbm h_1$ (component of $\mathbbm h$ in the $x_1$ direction) and the total Cauchy stress $\tau_{11}$ with time are plotted in Figs.~\ref{fig: no deformation vary_qv}--\ref{fig: no deformation vary_mag} for the following values of the magnetic induction.
\begin{equation}
\mathbb{B}_2 = \mathbb{B}_3 = 0, \quad
  \mathbb{B}_1 = \begin{cases}
    0, & \mbox{for}\;\; t<0,\\
    0.1 \, \mbox{T}, & \mbox{for}\; \; t\ge 0.
  \end{cases}
\end{equation}

We study the dependence of the magneto-viscoelastic coupling parameters $q_v$ and $r_v$, and the applied magnetic induction $\mathbb{B}_1$ on the relaxation of magnetic field and total Cauchy stress. 

It is seen from Fig.~\ref{fig: no deformation vary_qv}a that a large value of $q_v$ causes a high initial magnetic field but the decay to equilibrium value is also faster when $q_v$ is high. In the case of the total Cauchy stress, as seen from Fig.~\ref{fig: no deformation vary_qv}b, $q_v$ has no effect on the initial viscous overstress but a large $q_v$ also causes the stress to decay faster and reach the equilibrium value.  This is expected since the expression for $\mathbf{S}_v$ in Eq.~\eqref{Sv expression} does not contain $q_v$ explicitly but the dependence comes through the evolution equation~\eqref{Bv evolution equation}.
The parameter $r_v$ has a similar effect on the magnetic field as does $q_v$ but different in the case of the total Cauchy stress. As observed from Fig.~\ref{fig: no deformation vary_rv}, a large value of $r_v$ causes a higher initial stress and a faster decay of the same to equilibrium. The dependence of the relaxation processes on the applied magnetic induction is shown in Fig.~\ref{fig: no deformation vary_mag}. It is seen that the equilibrium values for all curves are different in this case since they depend on the value of applied magnetic induction. As expected from Eqs.~\eqref{taue expression} and \eqref{he expression} a higher magnetic induction causes a larger magnetic field and a larger stress. It is also observed that the higher the magnetic induction, the longer it takes for both the magnetic field and the stress to relax and reach equilibrium.

\subsection{Magnetic induction with a uniaxial deformation}
\begin{figure}
 \begin{center}
\includegraphics[scale=0.9]{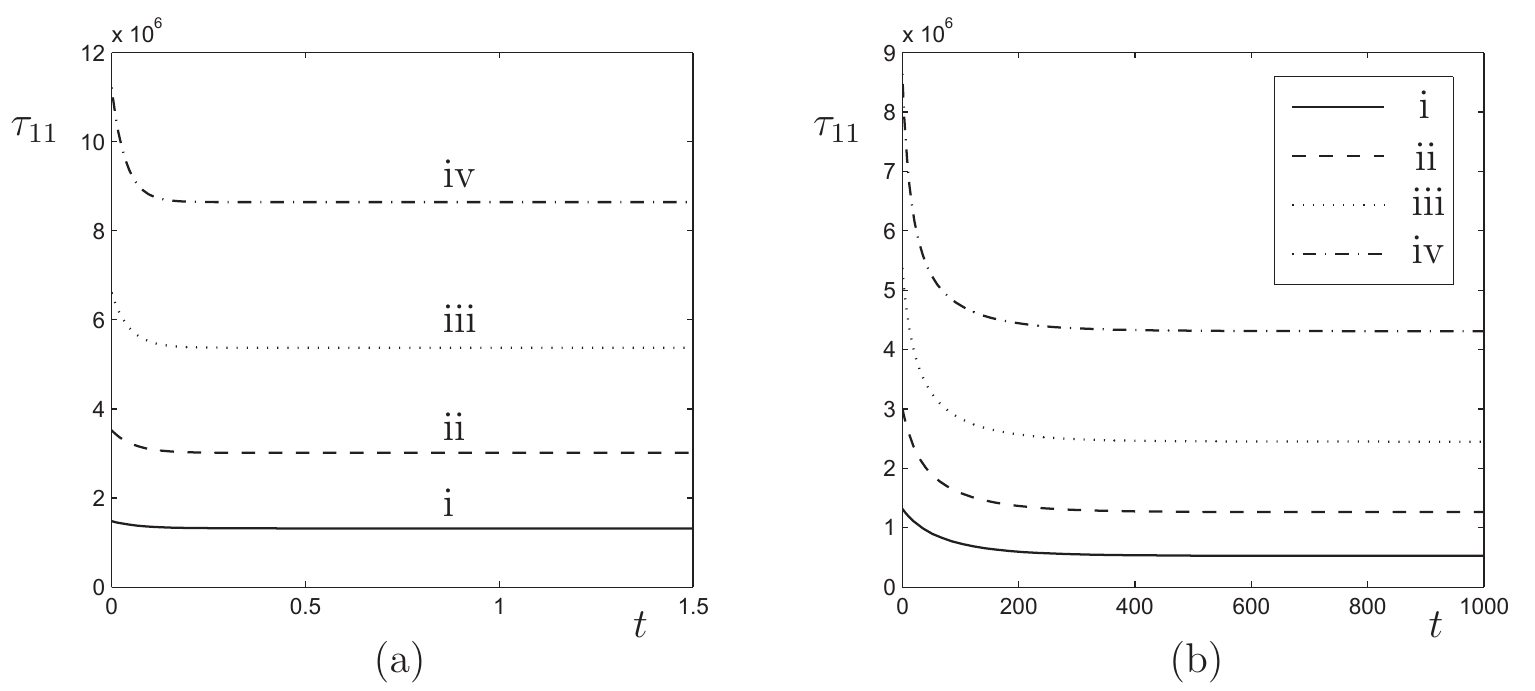}
 \end{center}
\caption{Uniaxial stretch in $x_1$ direction, $\mathbb B_1 = 0.1$ T. Variation of principal total Cauchy stress $\tau_{11}$ (N/m$^2$) vs time $t$ (s) for different values of the stretch $\lambda_1$. \textbf{(a)} small time scale \textbf{(b)} large time scale. (i) $\lambda_1=1.5$, (ii) $\lambda_1=2$, (iii) $\lambda_1=3$, (iv) $\lambda_1=4$.}
\label{fig: uniaxial -stress}
\end{figure}

In this case, we specify  a deformation and a magnetic induction in the $x_1$ direction while allowing the material to move freely in $x_2$ and $x_3$ directions. The stretch $\lambda_1$ and the magnetic induction $\mathbb B_1$ are applied at time $t=0$ and then the material is allowed to relax and reach an equilibrium state.
Variation of the magnetic field and the total Cauchy stress with time are plotted in Figs.~\ref{fig: uniaxial -stress} and \ref{fig: uniaxial - lam, Tv} for the values
\begin{equation}
  T_v = 100\, \mbox{s}, \quad \mathbb B_1 = 0.1\, \mbox{T}, \quad \lambda_1=1.5,
\end{equation}
unless otherwise stated to be different for individual problems.

\begin{figure}
 \begin{center}
\includegraphics[scale=0.9]{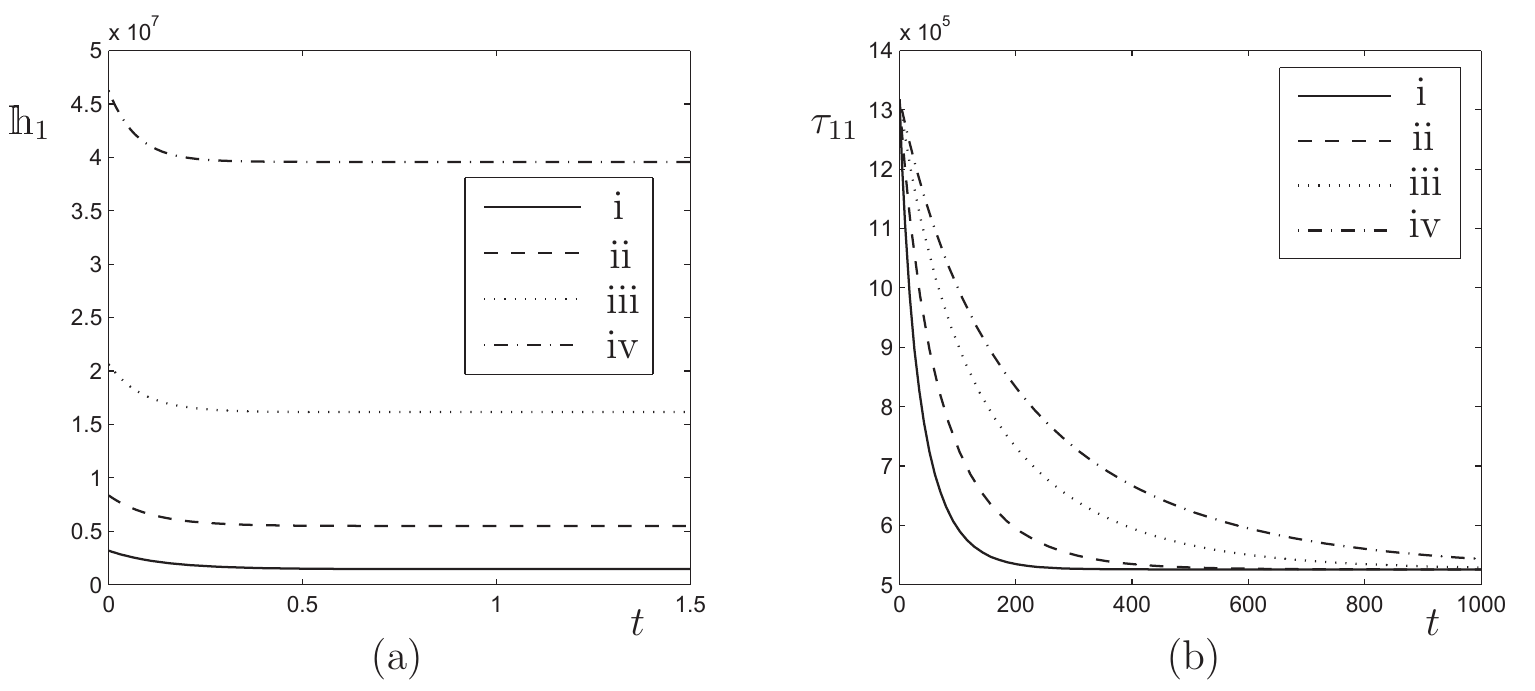}
 \end{center}
\caption{Uniaxial stretch in $x_1$ direction, $\mathbb B_1 = 0.1$ T. Variation of \textbf{(a)} total magnetic field $\mathbbm{h}_1$ (A/m) vs time $t$ (s) for different values of the stretch $\lambda_1$. (i) $\lambda_1=1.5$, (ii) $\lambda_1=2$, (iii) $\lambda_1=3$, (iv) $\lambda_1=4$; \textbf{(b)} principal total Cauchy stress $\tau_{11}$ (N/m$^2$) vs time $t$ (s) for different values of parameter $T_v$ (i) $T_v = 50$ s, (ii) $T_v = 100$ s, (iii) $T_v = 200$ s, (iv) $T_v = 300$ s.}
\label{fig: uniaxial - lam, Tv}
\end{figure}

It is observed from Fig.~\ref{fig: uniaxial -stress} that two different relaxations over different time scales occur in the total Cauchy stress $\tau_{11}$ -- one corresponding to the evolution of $\mathbb{B}_v$ and other corresponding to that of $\mathbf{C}_v$. 
The decay for small time scale (upto 1.5 seconds) to reach an equilibrium value is shown in Fig.~\ref{fig: uniaxial -stress}a while that for a longer time scale (upto 1000 seconds) is shown in Fig.~\ref{fig: uniaxial -stress}b. 
The four curves correspond to four different values of the initial stretch. It should be noted that the end point of a curve in Fig.~\ref{fig: uniaxial -stress}a is the same as the starting point of the corresponding curve in Fig.~\ref{fig: uniaxial -stress}b. 
A higher stretch causes an increase in the equilibrium value of the magnetic field in Fig.~\ref{fig: uniaxial - lam, Tv}a and also causes a faster relaxation to equilibrium value.
As expected from the existing results of pure mechanical viscoelasticity, cf. \cite{Hossain2012}, a higher stretch leads to a larger value of stress in Fig.~\ref{fig: uniaxial -stress} and a smaller value of $T_v$ causes stress to relax faster in Fig.~\ref{fig: uniaxial - lam, Tv}b.

\subsection{Time dependent deformation}
\begin{figure}
\begin{center}
\includegraphics[scale=0.9]{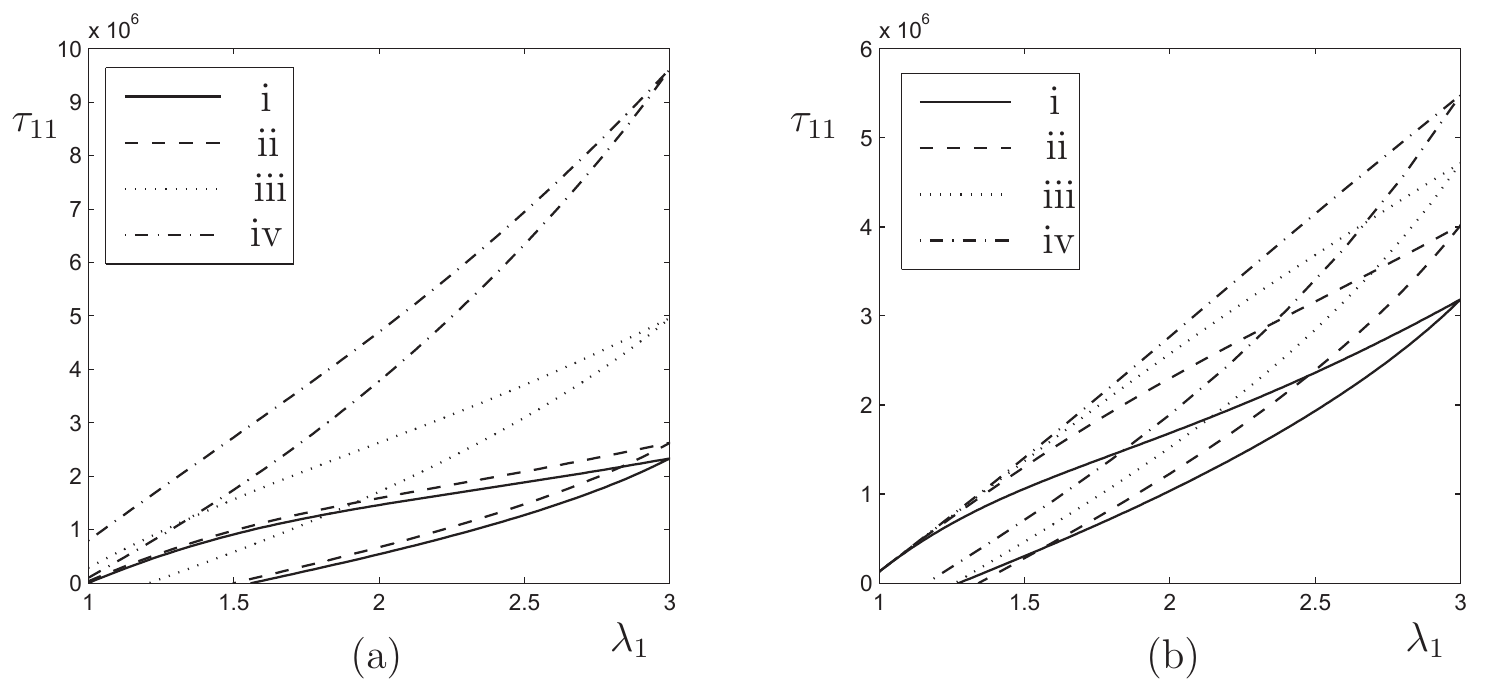}
\end{center}
\caption{Principal total Cauchy stress $\tau_{11}$ (N/m$^2$) vs stretch $\lambda_1$. \textbf{(a)}~Different values of the underlying magnetic induction (i)~$\mathbb B_1=0$, (ii)~$\mathbb B_1=0.1 $~T, (iii)~$\mathbb B_1=0.3$~T, (iv)~$\mathbb B_1=0.5$~T. \textbf{(b)}~Different values of the stretch rate (i)~$\dot{\lambda}_1= \pm 0.005 $~s$^{-1}$, (ii)~$\dot{\lambda}_1= \pm 0.02 $~s$^{-1}$, (iii)~$\dot{\lambda}_1= \pm 0.04 $~s$^{-1}$, (iv)~$\dot{\lambda}_1= \pm 0.08 $~s$^{-1}$.}
\label{fig: strain rate 1}
\end{figure}

\begin{figure}
\begin{center}
\includegraphics[scale=0.9]{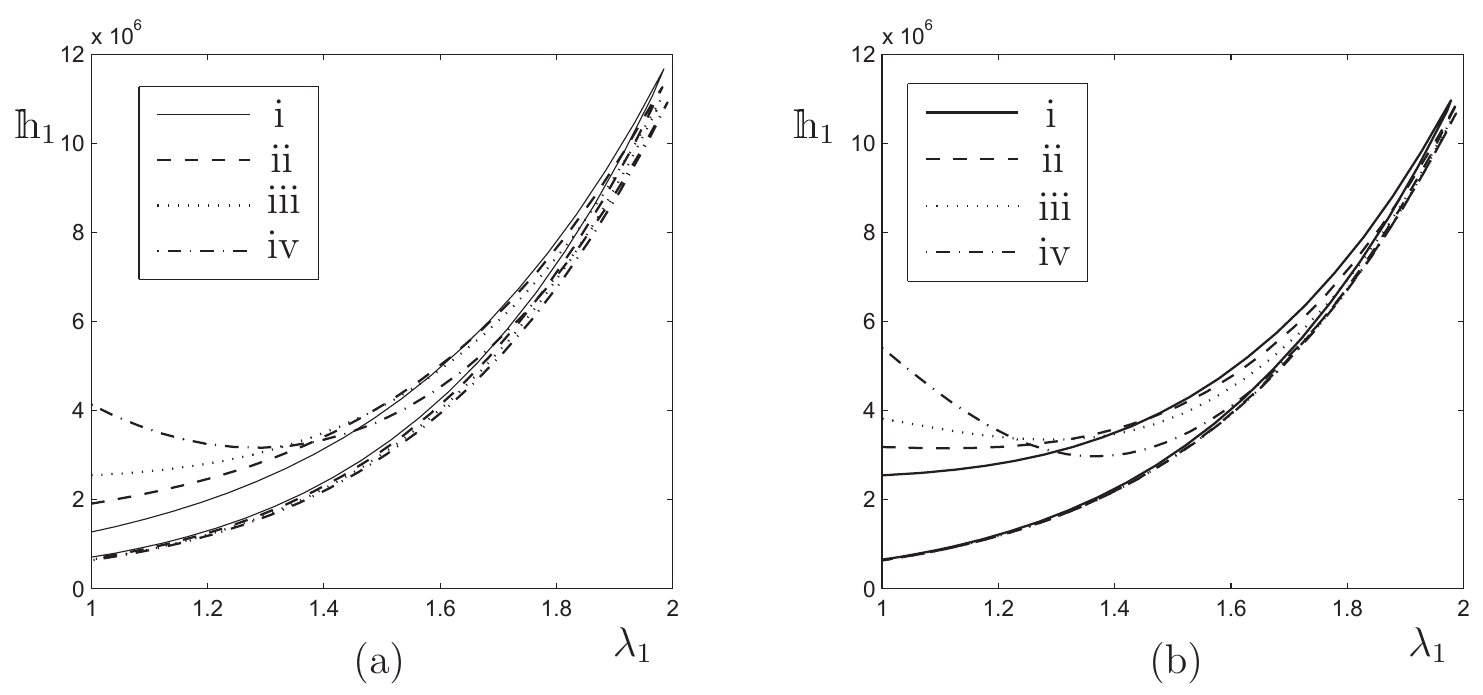}
\end{center}
\caption{Total magnetic field $\mathbbm{h}_{1}$ (A/m) vs stretch $\lambda_1$ at a stretch rate $\dot{\lambda}_1 = \pm 3 $ s$^{-1}$. \textbf{(a)}~Different values of the parameter $q_v$ (i)~$q_v = 1/\mu_0$,  (ii)~$q_v = 3/\mu_0$, (iii)~$q_v = 5/\mu_0$, (iv)~$q_v = 10/\mu_0$; \textbf{(b)}~Different values of the parameter $r_v$ (i)~$r_v = 1/\mu_0$,  (ii)~$r_v = 3/\mu_0$, (iii)~$r_v = 5/\mu_0$, (iv)~$r_v = 10/\mu_0$.}
\label{fig: strain rate 2}
\end{figure}

\begin{figure}
\begin{center}
\includegraphics[scale=0.9]{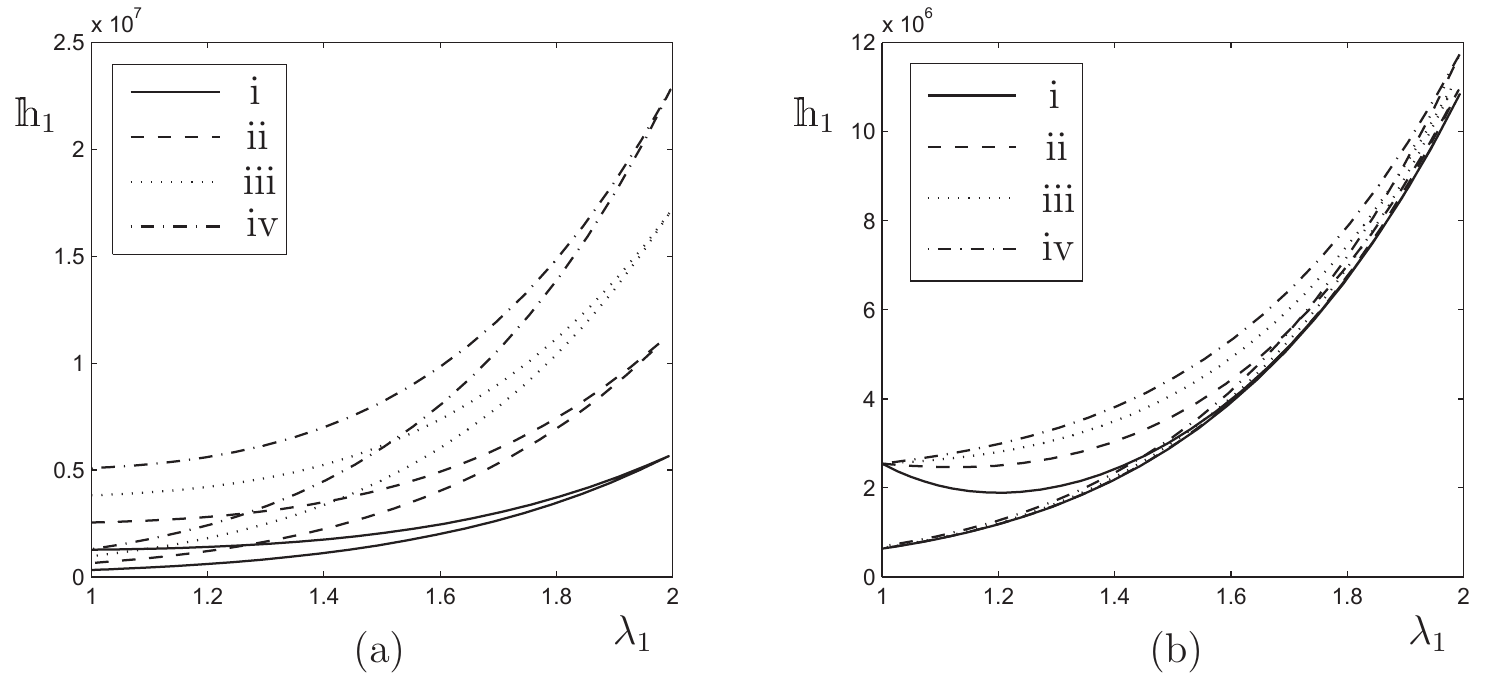}
\end{center}
\caption{Total magnetic field $\mathbbm{h}_{1}$ (A/m) vs stretch $\lambda_1$. \textbf{(a)}~Different values of the magnetic induction $\mathbb B_1$ at a stretch rate $\dot{\lambda}_1 = \pm 3 $ s$^{-1}$ (i)~$\mathbb B_1 = 0.1$ T,  (ii)~$\mathbb B_1 = 0.2$ T, (iii)~$\mathbb B_1 = 0.3$ T, (iv)~$\mathbb B_1 = 0.4$ T; \textbf{(b)}~Different values of the stretch rate $\dot{\lambda}_1$ (i)~$\dot{\lambda}_1 = 1$ s$^{-1}$,  (ii)~$\dot{\lambda}_1=2$ s$^{-1}$, (iii)~$\dot{\lambda}_1 = 3$ s$^{-1}$, (iv)~$\dot{\lambda}_1 = 4$ s$^{-1}$.}
\label{fig: strain rate 3}
\end{figure}

We now study the effects of the magneto-viscoelastic coupling on a dynamic deformation of the material.
In this case, calculations are performed corresponding to an experiment where a magnetic induction is applied at time $t=0$ and the material is stretched with a constant rate in the $x_1$ direction. On reaching $\lambda_1 = 3$, stretch is reduced at the same rate until a condition of zero stress or zero deformation (whichever earlier) is reached.
Effects on the total Cauchy stress and the total magnetic field of the applied magnetic induction, the rate of stretch, and the parameters $q_v$ and $r_v$ is analysed in Figs.~\ref{fig: strain rate 1}--\ref{fig: strain rate 3}.
The following values of the magnetic induction and the stretch rate are used
\begin{equation}
\mathbb{B}_1 = 0.2 \, \mbox{T}, \quad \dot{\lambda}_1 = \pm 0.01 \, \mbox{s}^{-1},
\end{equation}
unless otherwise stated to be different for individual calculations.

It is seen from Fig.~\ref{fig: strain rate 1}a that the starting points of all four curves are different corresponding to the stress induced due to the applied magnetic induction. The stress first increases with time (due to an increasing $\lambda_1$) and then falls with a decreasing $\lambda_1$ following a different path than earlier. A higher magnetic induction leads to larger value of the peak stress reached during the process. Similar curves for different values of the stretch rates are shown in Fig.~\ref{fig: strain rate 1}b. A larger value of stretch rate causes a larger peak value of stress since the material gets less time to relax as observed for the purely mechanical viscoelastic case by \cite{Lion1997} and \cite{Amin2006}.

Similar variation of the magnetic field $\mathbbm h_1$ with the stretch $\lambda_1$ can be observed for large values of stretch rates since $T_m$ is much smaller than $T_v$. The results for these calculations are shown in Figs.~\ref{fig: strain rate 2} and \ref{fig: strain rate 3} for $\lambda_1 = 2$ as the value of the maximum obtained stretch. As observed from Fig.~\ref{fig: strain rate 2}, starting at $t=0$, the magnetic field first falls and then rises due to an increase in  $\lambda_1$. As $\lambda_1$ reduces, $\mathbbm h_1$ comes down approaching a steady equilibrium value. High values of $q_v$ and $r_v$ cause a high initial magnetic field and a faster approach towards the equilibrium.

Dependence of this process on the applied magnetic induction and the stretch rate is shown in Fig.~\ref{fig: strain rate 3}. The different start and end points of the curves in Fig.~\ref{fig: strain rate 3}a correspond to the  values of magnetic field caused by different magnetic inductions. A higher magnetic induction causes a larger magnetic field while for a lower value of stretch rate, as observed from Fig.~\ref{fig: strain rate 3}b, the magnetic field approaches the steady equilibrium value earlier in the cycle.

\subsection{Time dependent magnetic induction}
\begin{figure}
 \begin{center}
\includegraphics[scale=0.9]{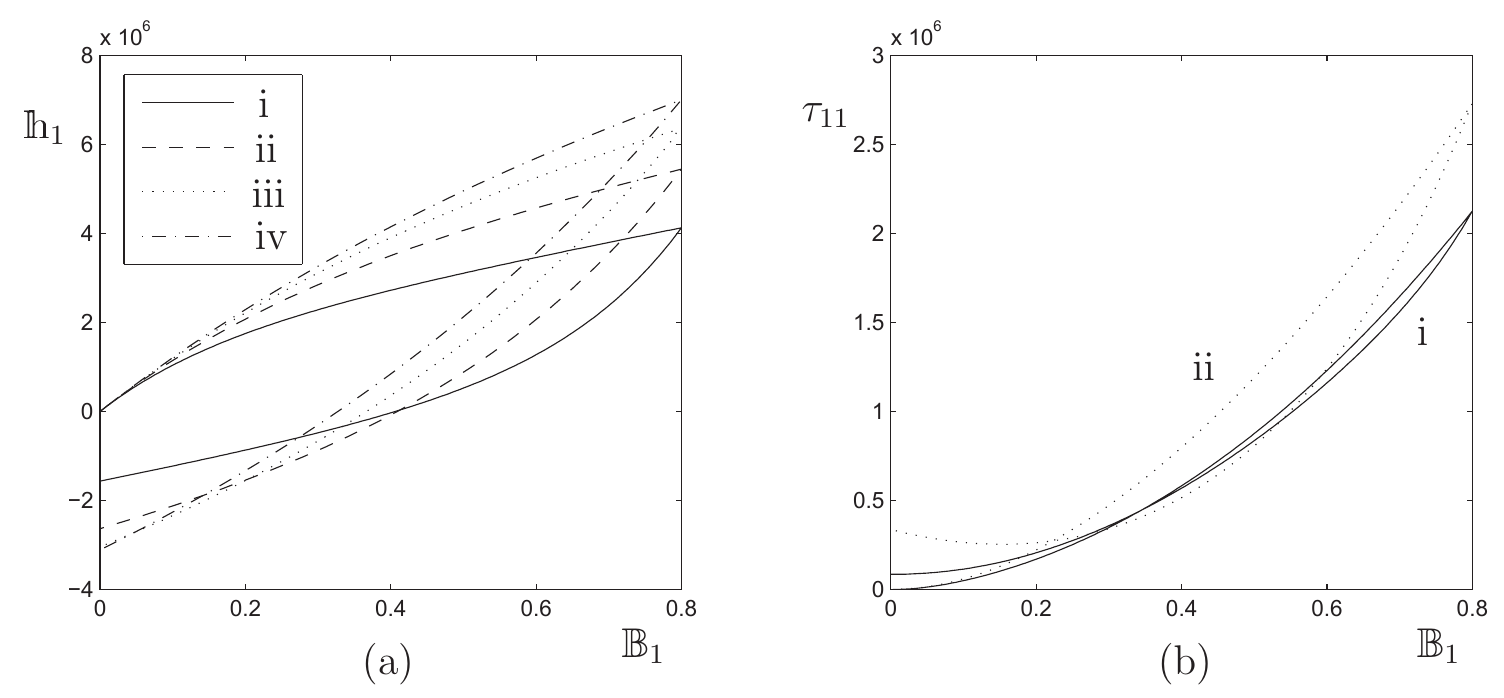}
 \end{center}
\caption{Time dependent magnetic induction, variation with the rate of magnetic induction \textbf{(a)}~Total magnetic field $\mathbbm{h}_{1}$~(A/m) vs magnetic induction $\mathbb{B}_1$~(T) (i)~$\dot{\mathbb B}_1= 1$~T/s,  (ii)~$\dot{\mathbb B}_1= 2$~T/s, (iii)~$\dot{\mathbb B}_1= 3$ T/s, (iv)~$\dot{\mathbb B}_1= 4$ T/s; \textbf{(b)}~Principal total cauchy stress $\tau_{11}$~(N/m$^2$) vs magnetic induction $\mathbb{B}_1$~(T). (i)~$\dot{\mathbb B}_1= 1$ T/s, (ii)~$\dot{\mathbb B}_1= 4$ T/s. }
\label{fig: mag rate 1}
\end{figure}

\begin{figure}
 \begin{center}
\includegraphics[scale=0.9]{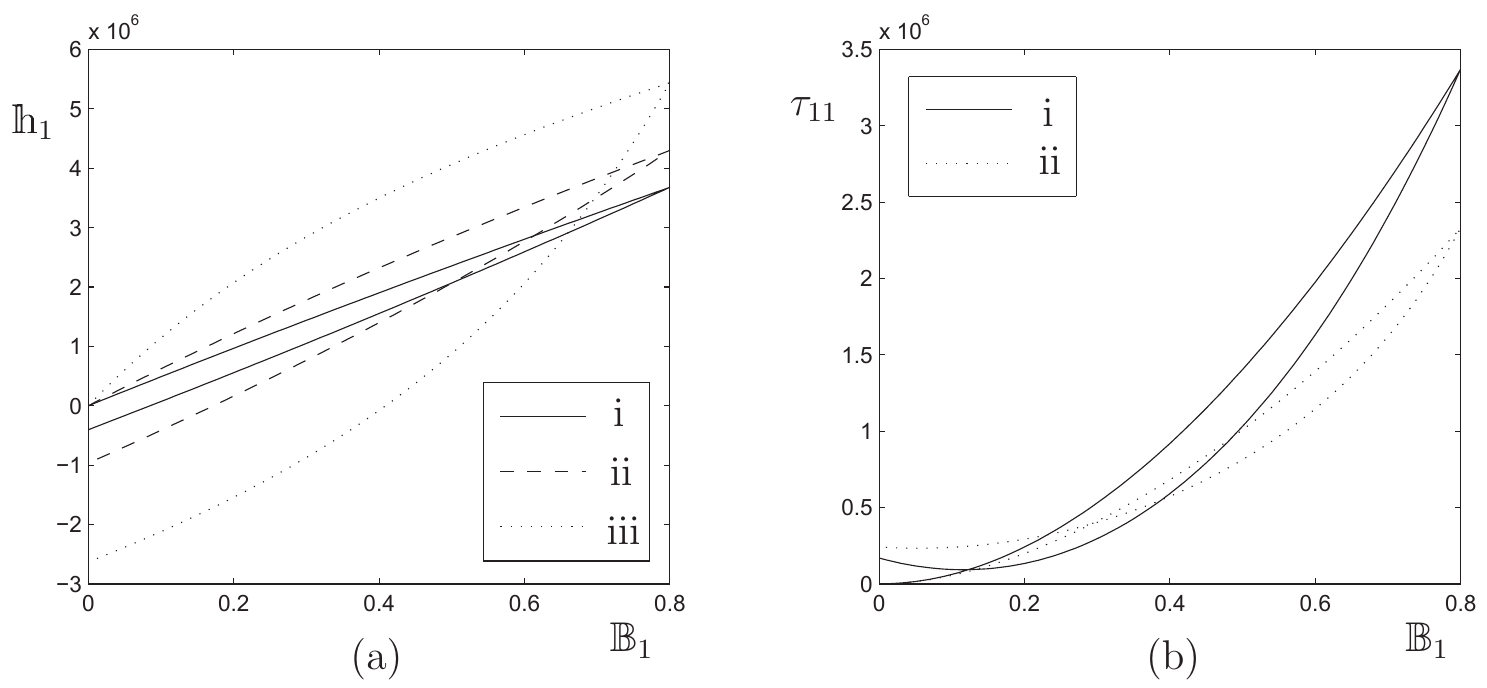}
 \end{center}
\caption{Time dependent magnetic induction, variation with the parameter $q_v$ \textbf{(a)}~Total magnetic field $\mathbbm{h}_{1}$~(A/m) vs magnetic induction $\mathbb{B}_1$~(T) (i)~$q_v = 0.1/\mu_0$, (ii)~$q_v = 1/\mu_0$, (iii)~$q_v = 5/\mu_0$; \textbf{(b)}~Principal total cauchy stress $\tau_{11}$~(N/m$^2$) vs magnetic induction $\mathbb{B}_1$~(T). (i)~$q_v = 0.1/\mu_0$, (ii)~$q_v = 5/\mu_0$. }
\label{fig: mag rate 2}
\end{figure}

\begin{figure}
 \begin{center}
\includegraphics[scale=0.9]{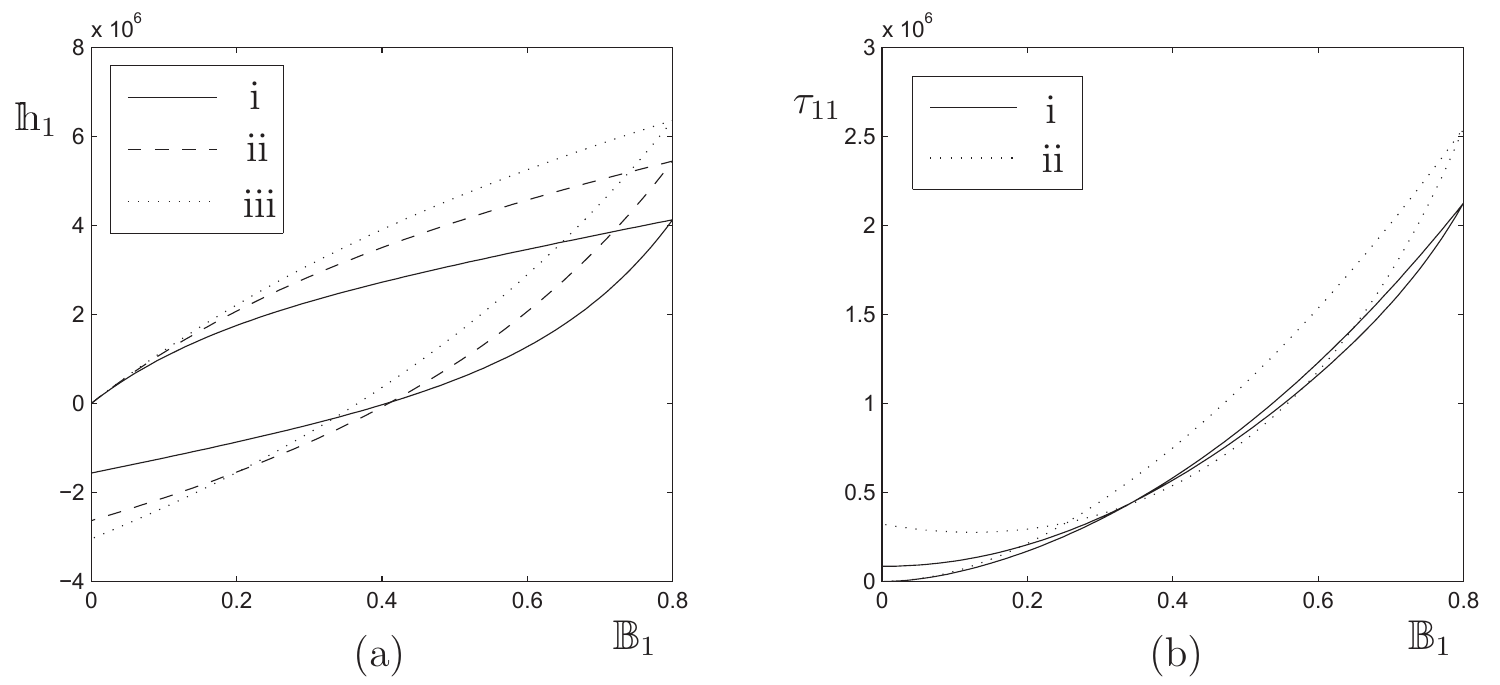}
 \end{center}
\caption{Time dependent magnetic induction, variation with the specific relaxation time $T_m$ \textbf{(a)}~Total magnetic field $\mathbbm{h}_{1}$ (A/m) vs magnetic induction $\mathbb{B}_1$ (T) (i)~$T_m=1$~s, (ii)~$T_m= 2$~s, (iii)~$T_m= 3$~s; \textbf{(b)}~Principal total cauchy stress $\tau_{11}$ (N/m$^2$) vs magnetic induction $\mathbb{B}_1$ (T). (i)~$T_m=1$~s, (ii)~$T_m = 3$~s. }
\label{fig: mag rate 3}
\end{figure}

\begin{figure}
\begin{center}
\includegraphics[scale=0.9]{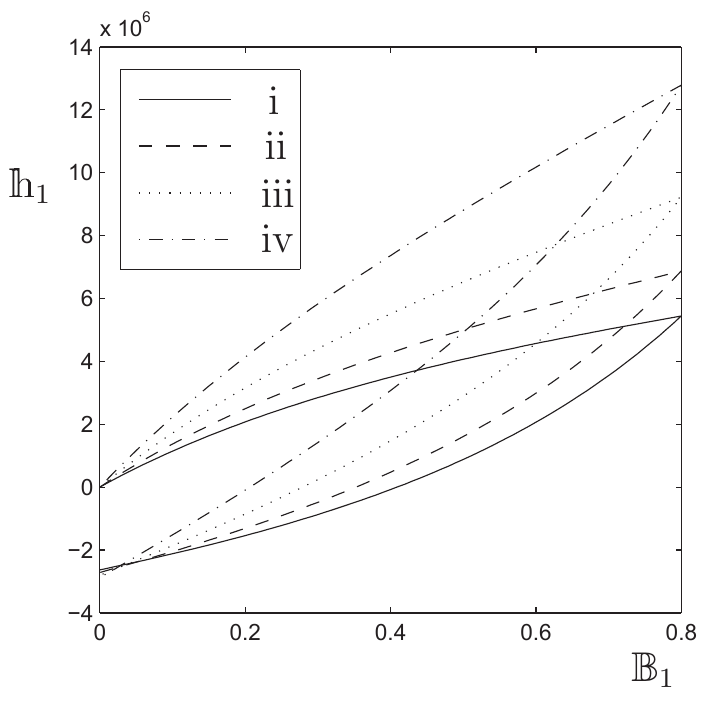}
\end{center}
\caption{Variation of the total magnetic field $\mathbbm{h}_1$~(A/m) with the magnetic induction $\mathbb B_1$~(T), $\dot{\mathbb B}_1 = 2$ T/s, (i)~$\lambda_1 = 1$, (ii)~$\lambda_1 = 1.2$, (iii)~$\lambda_1 = 1.4$, (iv)~$\lambda_1 = 1.6$.}
\label{fig: mag rate 4}
\end{figure}

In this case we study the effect of a time-varying magnetic induction on the induced stress and magnetic field in an undeformed material. The sample is assumed to be fixed at zero deformation and a magnetic induction is applied at time $t=0$ in the $x_1$ direction with a constant rate until a value of $B_1 = 0.8$ T is obtained. The induction is then reduced  with the same rate until it reaches zero.
Numerical results for this case are shown in Figs.~\ref{fig: mag rate 1}--\ref{fig: mag rate 4}. A value of $\dot{\mathbb B}_1 = 2$ T/s is used to plot the curves in Figs.~\ref{fig: mag rate 2}--\ref{fig: mag rate 4}.

It is observed from Fig.~\ref{fig: mag rate 1}a that for a particular value of induction rate, the magnetic field increases with an increasing magnetic induction and in the return cycle, it reduces to eventually obtain a negative value. In the entire cycle, the material develops a magnetisation in the $x_1$ direction due to the existing magnetic induction. As the induction reduces to zero, the material has to develop a magnetic field in the negative $x_1$ direction to erase the magnetisation in $x_1$ direction. Also, it can be seen by  Eq.~\eqref{three field relation eulerian} that if $\mathbbm b= \mathbf{0}$, then $\mathbbm h$ and $\mathbbm m$ obtain opposite signs.

A higher rate of magnetic induction causes the magnetic field to reach a high peak value and  it also causes the magnetic field to reach the maximum negative value when $\mathbb B$ vanishes completely. Stress in this case has a rather interesting variation as observed from Fig.~\ref{fig: mag rate 1}b. Starting from zero, the stress increases with an increasing $\mathbb B_1$ and then falls as $\mathbb B_1$ is reduced to zero. However, for a high induction rate of $\dot{\mathbb B}_1 = 4$ T/s, in the return cycle the stress approaches a minimum and then rises again.

A larger value of $q_v$ causes a higher value of peak magnetic field and a larger negative value at the end of cycle as $\mathbb B_1\rightarrow0$. Moreover, it also clearly causes larger energy dissipation during the cycle as the area inside the curve (iii) of Fig.~\ref{fig: mag rate 2}a is much larger than that in curve (i). The stress in Fig.~\ref{fig: mag rate 2}b has a higher peak value for a smaller value of $q_v$ while a larger value of $q_v$ causes  a higher stress at the end of the cycle.
A smaller value of the parameter $T_m$ helps the material to relax in lesser amount of time, hence the peak values of the magnetic field and stress reached in Fig.~\ref{fig: mag rate 3} in this case are lower.
A higher stretch in Fig.~\ref{fig: mag rate 4} causes a larger peak value of the magnetic field.

\section{Concluding remarks}
We have presented a theory to model nonlinear magneto-viscoelastic deformations in this paper. 
The deformation gradient is multiplicatively decomposed and the magnetic induction is additively decomposed to `elastic' and  `viscous' parts to take into account dissipation mechanisms. 
A Mooney--Rivlin type magnetoelastic energy density function is used for the equilibrium part, which is simplified to a neo-Hookean type energy density function for the non-equilibrium part of the free energy.
These, along with thermodynamically consistent evolution laws, are used to obtain numerical solutions corresonding to several different magneto-viscoelastic deformations.


The magneto-viscoelastic parameters $q_v$ and $r_v$ can have strong effects on the non-equilibrium magnetic field and the non-equilibrium total Cauchy stress by changing their peak values and the decay times. Strong couplings are also shown to exist between the magnetic induction and the non-equilibrium stress, and the underlying deformation and the non-equilibrium magnetic field, as is evident from Figs.~4b and 6a. We observe that a stretch rate and a magnetic induction rate can have a considerable influence on the total Cauchy stress and the magnetic field. The developed model seems to capture the magneto-viscoelastic phenomena quite nicely and on isolating mechanical viscoelastic effects, our results are qualitatively the same as those obtained earlier by  \cite{Amin2006} and \cite{Hossain2012}.
 
It should be noted that the numerical results presented here are representative solutions considering only one dissipation mechanism in the body. The theory can be easily generalised to include multiple mechanisms to match the experimental data. We have considered a specific type of phenomenologically motivated constitutive law for an isotropic material in this paper. The theory for an anistropic material and possibility of existence of other constitutive laws (such as those derived from micromechanics of the material) will be discussed in forthcoming contributions.

\bigskip
\noindent \textbf{Acknowledgements:} \\
This work is funded by an ERC advanced grant within the project MOCOPOLY.



\end{document}